\RequirePackage{fix-cm}
\documentclass[twocolumn,epjc3]{svjour3}  
\smartqed  
\RequirePackage{graphicx}

\RequirePackage{subfigure}
\RequirePackage{mathrsfs}
\RequirePackage{lineno}

\RequirePackage{cite}

\RequirePackage[colorlinks,breaklinks,pdftitle={ATLAS draft},pdfauthor={The ATLAS Collaboration}]{hyperref}
\hypersetup{linkcolor=blue,citecolor=blue,filecolor=black,urlcolor=blue}

\usepackage{preprintcover}  
\PreprintCoverPaperTitle{Measurement of the charge asymmetry in top quark \\ pair production in $pp$ collisions at
$\sqrt{s}=7$~TeV \\ using the ATLAS detector}  
\PreprintIdNumber{CERN-PH-EP-2012-015}  
\PreprintCoverAbstract{A measurement of the top-antitop production charge asymmetry \Ac\ is
  presented using data corresponding to an
  integrated luminosity of \lumi\ of $pp$ collisions at $\sqrt{s} =
  7$~TeV collected by the ATLAS detector at the LHC.  Events are selected
  with a single lepton (electron or muon), missing
  transverse momentum and at least four jets of which at least one jet is identified as coming
  from a $b$-quark.  A kinematic fit is
  used to reconstruct the \ttbar~event topology. 
  After background subtraction, a Bayesian unfolding procedure is performed to correct
  for acceptance and detector effects. The measured value of \Ac\ is \resultincl,
  consistent with the prediction from the \mcnlo\ Monte Carlo generator of $\Ac\ = 0.006\pm 0.002.$
  Measurements of \Ac\ in two ranges of invariant mass of the top-antitop pair are also shown.}  
\PreprintJournalName{Eur.\ Phys.\ J C}  

\newcommand{\ifb}{{\mbox{fb$^{-1}$}}}
\newcommand{\ttbar}{$t\bar{t}$}
\newcommand{\qqbar}{$q\bar{q}$}
\newcommand{\tbar}{\ensuremath{\bar{t}}}

\newcommand{\ET}{\ensuremath{E_{\mathrm{T}}}}
\def\TeV{\ifmmode {\mathrm{\ Te\kern -0.1em V}}\else
                   \textrm{Te\kern -0.1em V}\fi}%
\def\GeV{\ifmmode {\mathrm{\ Ge\kern -0.1em V}}\else
                   \textrm{Ge\kern -0.1em V}\fi}%
\def\MeV{\ifmmode {\mathrm{\ Me\kern -0.1em V}}\else
                   \textrm{Me\kern -0.1em V}\fi}%
\def\keV{\ifmmode {\mathrm{\ ke\kern -0.1em V}}\else
                   \textrm{ke\kern -0.1em V}\fi}%
\def\eV{\ifmmode  {\mathrm{\ e\kern -0.1em V}}\else
                   \textrm{e\kern -0.1em V}\fi}%

\newcommand{\met}{\ensuremath{E_{\mathrm{T}}^{\mathrm{miss}}}}

\newcommand{\pt}{\ensuremath{p_{\mathrm{T}}}}

\newcommand{\jimmy}{{\sc JIMMY}}
\newcommand{\herwig}{{\sc HERWIG}}
\newcommand{\powheg}{{\sc POWHEG}}
\newcommand{\pythia}{{\sc PYTHIA}}
\newcommand{\alpgen}{{\sc ALPGEN}}

\newcommand{\mcnlo}{{\sc MC@NLO}}
\newcommand{\sherpa}{{\sc SHERPA}}
\newcommand{\acermc}{{\sc ACERMC}}
\newcommand{\mttbar}{\ensuremath{m_{t\bar{t}}}}
\newcommand{\lumi}{1.04 \ifb}
\newcommand{\Yt}{\ensuremath{|y_t|}}
\newcommand{\Ytbar}{\ensuremath{|y_{\tbar}|}}
\newcommand{\DY}{\ensuremath{\Delta |y|}\ }
\newcommand{\ppbar}{\ensuremath{p\bar{p}}}
\newcommand{\Ac}{\ensuremath{A_C}} 
\newcommand{\resultincl}{\ensuremath{\Ac = -0.019 \pm 0.028 \ {\rm (stat.)} \pm 0.024 \ {\rm (syst.)}}}

\journalname{Eur. Phys. J. C}

\begin{document}

\modulolinenumbers[1]

\title{Measurement of the charge asymmetry in top quark pair production in $pp$ collisions at
$\sqrt{s}=7$~TeV using the ATLAS detector
}

\author{The ATLAS Collaboration\thanksref{e1,addr1}\\
}

\thankstext{e1}{e-mail: atlas.publications@cern.ch}


\institute{CERN, 1211 Geneva 23, Switzerland \label{addr1}
}

\date{Received: date / Accepted: date}

\maketitle

\begin{abstract}
A measurement of the top-antitop production charge asymmetry \Ac\ is
  presented using data corresponding to an
  integrated luminosity of \lumi\ of $pp$ collisions at $\sqrt{s} =
  7$~TeV collected by the ATLAS detector at the LHC.  Events are selected
  with a single lepton (electron or muon), missing
  transverse momentum and at least four jets of which at least one jet is identified as coming
  from a $b$-quark.  A kinematic fit is
  used to reconstruct the \ttbar~event topology. 
  After background subtraction, a Bayesian unfolding procedure is performed to correct
  for acceptance and detector effects. The measured value of \Ac\ is \resultincl,
  consistent with the prediction from the \mcnlo\ Monte Carlo generator of $\Ac\ = 0.006\pm 0.002.$
  Measurements of \Ac\ in two ranges of invariant mass of the top-antitop pair are also shown.
\end{abstract}
\section{Introduction}
\label{Introduction}
The top quark is the heaviest elementary particle so far observed. 
With a mass close to the electroweak scale it may play a special role in physics beyond the Standard Model (SM).
Its pair production at hadron colliders allows a test of quantum chromodynamics (QCD) at high energies.

This paper describes the measurement of the charge asymmetry $A_C$, defined as \cite{Jung:2011zv,Diener:2009ee}: 
\begin{equation}
\Ac = \frac{N(\Delta |y| >0) - N(\Delta |y| <0)}{N(\Delta |y| >0) + N(\Delta |y| <0)},
\label{Ac}
\end{equation}
where $\DY \equiv \Yt - \Ytbar$  is the difference between the absolute values of the top and antitop rapidities (\Yt\ and \Ytbar)
and $N$ is the number of events with \DY positive or negative.

Although \ttbar\ production at hadron colliders is predicted to be symmetric under the exchange of $t$ and $\bar{t}$ at leading order, 
at next-to-leading order (NLO) the process $q\bar{q}\to t\bar{t}g$ exhibits an asymmetry in the
differential distributions of the top and antitop, due to interference between initial and final state gluon emission. 
The $q\bar{q}\to t\bar{t}$ process also possesses an asymmetry due to the interference between the Born and box diagrams. 
Similarly, the $qg\to t\bar{t}q$ process is asymmetric due to interference between amplitudes which have a relative
sign difference under the exchange of $t$ and $\bar{t}.$ The production of \ttbar\ pairs by gluon-gluon fusion, $gg \to t\bar{t}$, on 
the other hand, is symmetric.

In \ppbar\ collisions at the Tevatron, where top pairs are predominantly produced by quark-antiquark annihilation, 
perturbative QCD predicts that the top quark will be preferentially emitted in the direction of the incoming quark and 
the antitop in the direction of the incoming antiquark~\cite{Kuhn:1998jr}.
Consequently, the charge asymmetry is measured as a forward-backward asymmetry, $A_{FB}.$
Recent measurements of $A_{FB}$ by the CDF and D0 Collaborations~\cite{D02,CDF1,CDF2,D01} show a $2$-$3 \sigma$ excess over the SM expectations enhancing interest in scrutinising the \ttbar\ asymmetry.
For \ttbar\ invariant mass, $\mttbar$, greater than $450\,\rm{GeV}$, the 
CDF experiment measures 
an asymmetry in the \ttbar\ rest frame which is 3.4$\sigma$ above the SM prediction~\cite{CDF2}. 
Several new physics models have been proposed to explain the excess 
observed at CDF and D0~\cite{Ferrario:2009bz,Frampton:2009rk,Jung:2009jz,Cheung:2009ch,Shu:2009xf,Dorsner:2009mq,Grinstein:2011yv,Ligeti:2011vt,Jung:2011zv,AguilarSaavedra:2011vw,AguilarSaavedra:2011hz}. 
Different models predict different asymmetries as a function of $\mttbar$~\cite{AguilarSaavedra:2011ci}.

In $pp$ collisions at the LHC, the dominant mechanism for \ttbar\ production is expected to be the gluon-gluon 
fusion process, while \ttbar\ production via \qqbar\ or $qg$ is small. 
Since the initial state is symmetric, the forward-backward asymmetry is no longer a useful observable.
However, due to the asymmetry in the production via $q\bar{q}$ and $qg$, QCD predicts at the LHC a small excess of centrally produced antitop quarks while top quarks are produced, 
on average, at higher absolute rapidities.
This can be understood by the fact that for \ttbar\ production via \qqbar\ annihilation
the valence quark carries, on average, a larger momentum fraction than the anti-quark from the sea.
With top quarks preferentially emitted in the direction of the initial quarks in the \ttbar\ rest frame, the boost into 
the laboratory frame drives the top mainly in the forward or backward directions, while antitops are preferentially retained in the central region.
If new physics is responsible for the Tevatron $A_{FB}$ excess, the charge asymmetry measured at the LHC is a natural place to look for it.

In this paper, the measurement of the charge asymmetry $A_C$ is performed using candidate \ttbar\ events selected in the lepton+jets channel. In this channel, 
the SM decay of the \ttbar\ pair to $W^+ b W^- \bar{b}$ results in a single electron or muon from one of the $W$ boson decays and four jets, two from the second $W$ boson
decay and two from the $b$- and $\bar{b}$-quarks. 
To allow comparisons with theory calculations, the measured \DY distribution is unfolded to account for acceptance and detector effects.
An inclusive measurement, and measurements of \Ac\ in two ranges of \ttbar\ invariant mass, are presented.
An inclusive measurement of this asymmetry with an equivalent observable has been recently reported by the CMS collaboration~\cite{Chatrchyan:2011hk}.

\section{The ATLAS detector}
\label{Detector}
The ATLAS detector~\cite{Aad:2008zzm} at the LHC covers nearly the entire
solid angle\footnote{ATLAS uses a right-handed coordinate system with 
its origin at the nominal interaction point (IP) in the centre of the detector and the $z$-axis along the beam pipe. 
The $x$-axis points from the IP to the centre of the LHC ring, and the $y$ axis points upward. Cylindrical 
coordinates $(r,\phi)$ are used in the transverse plane, $\phi$ being the azimuthal angle around the beam pipe. 
The pseudorapidity is defined in terms of the polar angle $\theta$ as $\eta=-\ln\tan(\theta/2)$.
Transverse momentum and energy are defined as $\pt = p\sin\theta$ and
$\ET = E\sin\theta$, respectively.} around the collision point. It consists of an inner 
tracking detector surrounded by a thin 
superconducting solenoid, electromagnetic and hadronic calorimeters,
and an external muon spectrometer incorporating three large
superconducting toroid magnet assemblies. 

The inner-detector system is immersed in a 2\,T axial magnetic field
and provides charged particle tracking in the range $|\eta|<2.5$.  The
high-granularity silicon pixel detector covers the vertex region and
provides typically three measurements per track, followed by the
silicon microstrip tracker (SCT) which provides four measurements from
eight strip layers. These silicon detectors are complemented by the
transition radiation tracker (TRT), which enables extended track
reconstruction up to $|\eta|=2.0$. In giving typically more than 30
straw-tube measurements per track, the TRT improves the inner
detector momentum resolution, and also provides electron
identification information.

The calorimeter system covers the pseudorapidity range $|\eta|<4.9$.
Within the region $|\eta|<3.2$, electromagnetic calorimetry is
provided by barrel and endcap lead-liquid argon (LAr) electromagnetic
calorimeters, with an additional thin LAr presampler covering
$|\eta|<1.8$ to correct for energy loss in material upstream of the
calorimeters. Hadronic calorimetry is provided by the
steel/scintillating-tile calorimeter, segmented into three barrel
structures within $|\eta|<1.7$, and two copper/LAr hadronic endcap
calorimeters. The solid angle coverage is completed with forward
copper/LAr and tungsten/LAr calorimeter modules optimised for
electromagnetic and hadronic measurements respectively.

The muon spectrometer comprises separate trigger and high-precision
tracking chambers measuring the deflection of muons in a magnetic
field with a bending integral from 2 to 8 Tm in the central region,
generated by three superconducting air-core toroids. The precision
chamber system covers the region $|\eta|<2.7$ with three layers of
monitored drift tubes, complemented by cathode strip chambers in the
forward region, where the background is highest. The muon trigger
system covers the range $|\eta|<2.4$ with resistive plate chambers in
the barrel, and thin gap chambers in the endcap regions.

A three-level trigger system is used to select interesting events. The
\mbox{level-1} trigger is implemented in hardware and uses a subset of
detector information to reduce the event rate to a design value of at most
75\,kHz. This is followed by two software-based trigger levels,
\mbox{level-2} and the event filter, which together reduce the event rate to
about 300\,Hz.

\section{Data and Monte Carlo samples}
\label{Samples}
Data from LHC $pp$ collisions collected by the ATLAS detector between March and June 2011 are used in the analysis, corresponding to an integrated luminosity of \lumi.

Simulated top pair events are generated using the \mcnlo\cite{FRI-0201} Monte Carlo (MC) generator with the NLO
parton density function (PDF) set CTEQ6.6~\cite{Nadolsky:2008zw}.
Parton showering and the underlying event are modelled using \herwig~\cite{COR-0001} and 
\jimmy~\cite{JButterworth:1996zw}, respectively.
This \ttbar\ sample is normalised to a cross section of 165 pb, obtained with 
the latest theoretical computation, which approximates the 
next-to-next-to leading order prediction~\cite{Aliev:2010zk}.
Single top events are also generated using \mcnlo \ while the production of $W/Z$ bosons in association with jets is simulated using 
the \alpgen \ generator~\cite{MAN-0301} interfaced to \herwig \ and \jimmy\ with CTEQ6.1~\cite{cteq6}.
Diboson events ($WW$, $WZ$, $ZZ$) are generated using \herwig\ with MRST2007lomod~\cite{Sherstnev:2007nd}. 

All Monte Carlo simulation samples are generated with multiple $pp$ interactions per bunch crossing (pile-up).
These simulated events are re-weighted so that the distribution of the number of interactions per crossing 
in simulation matches that in the data.
The samples are then processed through the GEANT4~\cite{AGO-0301} 
simulation~\cite{2010EPJC...70..823A} of the ATLAS detector and the standard reconstruction software.

\section{Event selection}
\label{Selection}
\subsection{Physics object selection}
Reconstructing top quark pair events in the detector requires electrons, muons, jets and missing momentum to be simultaneously measured. Electron candidates are 
defined as energy deposits in the electromagnetic calorimeter associated with a well-measured track. 
Identification criteria based on shower shape variables, track quality, and information from the transition radiation tracker are applied to electron candidates~\cite{Aad:2011mk}.
All candidates are required to have $\pt>25$ GeV and $|\eta_{\rm cluster}|<2.47$,
where $\eta_{\rm cluster}$ is the pseudorapidity of the electromagnetic calorimeter cluster associated with the electron. Candidates in the calorimeter transition region $1.37<|\eta_{\rm cluster}|<1.52$ are excluded. 

Muon candidates are reconstructed from track segments in different layers of the muon chambers. These segments are
combined starting from the outermost layer, with a procedure that takes material effects into account,
and matched with tracks found in the inner detector. The candidates are then refitted 
using the complete track information from both detector systems, and are required to satisfy $\pt>20$ GeV and $|\eta|<2.5$. 

Jets are reconstructed with the anti-$k_t$ algorithm, with a distance parameter of $0.4$~\cite{Cacciari:2008gp}, starting 
from clusters of energy in adjacent calorimeter cells at the electromagnetic (EM) scale.
The jet energy is corrected to the hadronic scale
using $\pt$- and $\eta$-dependent correction factors obtained from simulation and validated with data~\cite{Aad:2011he}.
Jet quality criteria are applied to identify jets
not associated to in-time real energy deposits in the calorimeters caused by
various sources (calorimeter noise, non-collision beam-related background, cosmic-ray induced showers).

The missing transverse momentum (\met) is reconstructed from clusters of energy calibrated at the EM scale
and corrected according to the energy scale of the associated physics object~\cite{Aad:2012re}. Contributions from
muons are included using their momentum measured from the tracking and muon spectrometer systems.
The remaining clusters not associated with the high \pt\ objects are also included in the missing transverse momentum.

Muons within $\Delta R=0.4$ of a jet axis\footnote{
$\Delta R = \sqrt{{\Delta \phi}^2+{\Delta \eta}^2}$, where $\Delta\phi$ and $\Delta\eta$
are the separation in azimuthal angle and pseudorapidity, respectively.} and with $\pt>20$ GeV are removed in order to reduce the 
contamination caused by muons from hadron decays. Subsequently, jets within $\Delta R=0.2$ of an electron candidate are removed 
to avoid double counting electrons as jets.

Isolation criteria are applied to both electron and muon candidates to 
reduce the backgrounds from hadrons mimicking lepton signatures and backgrounds 
from heavy flavour decays inside jets. For electrons, the 
total energy in a cone of $\Delta R=0.2$ around the electron candidate 
must not exceed $3.5$ GeV, after correcting for energy deposits from 
pile-up and for the energy associated with the electron. For 
muons, the sum of track transverse momenta for all tracks with $\pt>1$ GeV 
and the total energy deposited in a cone of $\Delta R=0.3$ around the muon are both required to be less 
than 4 GeV ignoring the contribution of the muon \pt.

Reconstructing top quark pair events is facilitated by the ability to 
tag jets from the hadronisation of $b$-quarks. For this purpose, 
two $b$-tagging algorithms are used and their results are combined to extract a tagging decision for each jet. 
One $b$-tagger exploits the topology of $b$- and $c$-hadron weak decays inside the jet. 
A Kalman filter~\cite{Kalman_1960} is used to find a common line on which the primary vertex and the $b$- and $c$-hadron 
decay vertices lie, as well as their position on this line, giving an approximate flight path 
for the $b$- and $c$-hadrons. 
The discrimination between $b$-, $c$-
and light quark jets is based on a likelihood using the masses, momenta, flight-length significances, 
and track multiplicities of the reconstructed vertices as inputs. To further increase the 
flavour discrimination power, a second $b$-tagger is run which does not attempt to directly 
reconstruct decay vertices. Instead, this second tagger uses the transverse and the 
longitudinal impact parameter significances of each track within the jet to determine a 
likelihood  that the jet originates from a $b$-quark. The results of both taggers are 
combined using a neural network to determine a single discriminant variable 
which is used to make tagging decisions. 
The combined tagger operating point chosen for the present analysis corresponds to a 
70\% tagging efficiency for $b$-jets in simulated $t\bar{t}$ events
while light flavor jets are suppressed by approximately a factor of 100.

\subsection{Selection of \ttbar\ candidates}
\label{evselection}


The \ttbar\ final state in the lepton+jets channel is characterised by an isolated
lepton (electron or muon) with relatively high \pt, missing transverse momentum arising
from the neutrino from the leptonic $W$ decay, two $b$-quark jets and
two light quark jets from the hadronic $W$ decay.  
To select events with this topology, the appropriate single-electron or single-muon trigger 
is required to have fired (with thresholds at 20 and 18~GeV respectively).
The events are also required to contain one and only one reconstructed lepton 
with $\pt > 25 \GeV$ for electrons and $\pt > 20 \GeV$ for muons. 
To reject multijet background
in the muon channel, $\met>20\GeV$ and $\met+m_{\rm T}(W)>60\GeV$ are required\footnote{Here $m_{\rm T}(W)$ is the $W$-boson transverse mass, defined as
$\sqrt{ 2 \pt^{\ell} \pt^{\nu} (1-\cos ( \phi^{\ell} - \phi^{\nu} )) }$
where the measured $\met$ vector provides the neutrino information.}.
In the electron channel more stringent cuts on \met\ and $m_{\rm T}(W)$ are required because of the higher level of multijet
background, i.e. $\met > 35\GeV$ and $m_{\rm T}(W)>25\GeV$. 
Events are required to have at least four jets with $\pt>25\GeV$ and $|\eta|<2.5$. 
These requirements define the `pretag' selection. The `tagged' selection requires, in addition, at least one
of the jets with  $\pt>25\GeV$ and $|\eta|<2.5$ to be $b$-tagged.

\section{Background determination}
\label{Backgrounds}
\subsection{Multijet background}
\label{fakes}
The method used for evaluating the multijet background with fake leptons\footnote{The term `fake' leptons here refers 
to hadrons mimicking lepton signatures and to leptons arising from heavy hadron decays, whereas `real' leptons come from $W$ and $Z$ decays.}
in both the electron and muon channels is the so-called `Matrix Method'.
This relies on defining loose and tight lepton samples~\cite{Aad:2010ey}
and measuring the fractions of real ($\epsilon_\mathrm{real}$) and fake ($\epsilon_\mathrm{fake}$) loose
leptons that are selected as tight leptons.
The fraction $\epsilon_\mathrm{real}$ is measured using data control samples of $Z$ boson decays to two leptons,
while  $\epsilon_\mathrm{fake}$ is measured from data control regions defined separately for the electron and muon
channels, where the contribution of fake leptons is dominant.

For the muon channel, the loose data sample is defined by removing the isolation requirements in the default
muon selection. The fake lepton efficiencies are determined using a low $m_{\rm T}$ control region $m_{\rm T} < 20\,\GeV$
with an additional cut $\met + m_{\rm T} < 60\,\GeV$.
The efficiencies for signal and fake leptons are parameterised as a function of muon $|\eta|$ and \pt\ in order to account for 
the variation of the muon detector acceptance and the profile of hadronic activity in the detector that affects the muon isolation. 

For the multijet background estimate in the electron channel, the loose
data sample is defined by considering events with electrons passing looser identification criteria. The electron isolation
requirement is also modified: the total energy in a cone of $\Delta R=0.2$ around the
electron is required to be smaller than $6$ \GeV\ (instead of 3.5 \GeV),
after correcting for energy deposits from pile-up interactions and for the energy associated with the electron. 
The fake lepton efficiencies are determined using a low $\met$ control region
($5\, \GeV < \met < 20\, \GeV$).

In both channels contributions from $W$+jets and $Z$+jets backgrounds in the control region, estimated
using Monte Carlo simulation, are subtracted.

\subsection{$W$+jets background estimation}
\label{sec:ChAsymm}

At the LHC the rate of $W^{+}$+jets is larger than that of $W^{-}$+jets
because there are more valence $u$ quarks than $d$ quarks in the proton.
Theoretically, the ratio of $W^+$+jets and $W^-$+jets cross sections is predicted much more precisely
than the total $W$+jets cross section \cite{Martin:2009iq,Kom:2010mv}. This asymmetry is exploited here to measure 
the total $W$+jets background from the data.

Since, to a good approximation, processes other than $W$+jets 
give equal numbers of positively and negatively charged leptons, the formula
\begin{equation}
\label{chargeaform}
N_{W^+} + N_{W^-} = \left( {r_{MC}+1 \over r_{MC}-1} \right) (D^+ - D^-),
\end{equation}
can be used to estimate the total number of $W$ events in the selected sample. Here $D^+ (D^-)$
are the total numbers of events in data passing the selection cuts described in Section~\ref{evselection} (apart from the $b$-tagging requirement)
with positively (negatively) charged leptons, and $r_{MC} \equiv {N(pp\rightarrow W^+ ) \over 
N(pp\rightarrow W^- )}$ is evaluated from Monte Carlo simulation, using the same event selection. 

The ratio $r_{MC}$ is found to be $1.56 \pm 0.06$ in the electron channel and $1.65\pm 0.08$ in the muon
channel. The dominant uncertainties on $r_{MC}$ originate from
those of the parton distribution functions, the jet energy scale,
and the heavy flavour fractions in $W$+jets events
(fractions of $W$+jets events containing $b\bar{b}$ pairs, $c\bar{c}$ pairs and $c$ quarks).

Since the theoretical prediction for heavy flavour fractions 
in $W$+jets suffers from large uncertainties, a data-driven
approach was developed to constrain these fractions with some inputs from MC simulation. 
In this approach
samples with a lower jet multiplicity, obtained from the selection
described in Section~\ref{evselection}, but requiring precisely one or
two jets instead of four or more jets, are analysed.  The numbers $W^{\rm Data}_{\rm i, pretag}, W^{\rm Data}_{\rm i, tagged}$, 
of $W+i$ jet events in these samples (where $i=1,2$), before and after applying the $b$-tagging requirement, are
computed by subtracting the small contributions of other Standard
Model processes - electroweak ($WW$, $WZ$, $ZZ$ and $Z$+jets) and top
($t\bar{t}$ and single top) using predictions from the simulation, and
by subtracting the multijet background as described in Section~\ref{fakes}.

A system of two equations, expressing the number of $W+1$~jet
events and $W+2$~jets events before and after $b$-tagging, can be written
with six independent flavour fractions as the unknowns, corresponding to fractions of
$Wb\bar{b}$+jets, $Wc\bar{c}$+jets, and $Wc$+jets events in the one and two jet bins. The simulation prediction
for the ratio of the heavy flavour fractions between the one and two jet bins is used
to relate the heavy flavour fractions in the two bins, reducing the number of independent
fractions to three. Finally, the ratio of the fractions of $Wc\bar{c}$+jets and
$Wb\bar{b}$+jets events in the two-jet bin is taken to be fixed to the value obtained from simulated
events in order to obtain two equations for two independent fractions.
Based on this measurement, the heavy flavour fractions in simulated
$W$+jets events are adjusted by a scale factor $1.63\pm 0.76$ for
$Wb\bar{b}$+jets and $Wc\bar{c}$+jets events and $1.11\pm 0.35$ for
$Wc$+jets. When applied to the signal region, an additional
25\% uncertainty on these fractions is added, corresponding to the
uncertainty of the Monte Carlo prediction for the ratio of heavy flavour
fractions in different jet multiplicities. The heavy flavour scale
factors are applied to simulated $W$+jets events throughout this
paper, and the effect of their uncertainties on the value of $r_{MC}$
is evaluated.

Using Equation (\ref{chargeaform}), the total number of $W$+jets events passing the event selection
described in Section~\ref{evselection} without requiring a $b$-tagged jet, $W_{\ge4,\rm pretag}$, is evaluated to be 
$5400 \pm 800\ ({\rm stat.+syst.})$ in the electron channel and $8600 \pm 1200\ ({\rm
stat.+syst.})$ in the muon channel. 

The number of $W$+jets events passing
the selection with at least one $b$-tagged jet is subsequently evaluated as~\cite{Aad:2010ey}
\begin{equation}
  \label{eq:taggedfrac}
  W_{\ge4, \rm tagged} = W_{\ge4, \rm pretag} \cdot f_{2,\rm tagged} \cdot k_{2\to\ge4}.
\end{equation}
\noindent 
Here $f_{\rm 2,tagged}\equiv W^{\rm Data}_{\rm 2, tagged}/$$W^{\rm Data}_{\rm 2, pretag}$
is the fraction of $W$+2~jets events passing the requirement of having at least one $b$-tagged jet, and
$k_{2\to\ge4}\equiv f^{\rm MC}_{\ge4\rm ,tagged}/$$f^{\rm MC}_{\rm 2,tagged}$ is the ratio of the fractions of simulated $W$+jets events passing the 
requirement of at least one $b$-tagged jet, for at least four and two jets, respectively. 
The value of $f_{2,\rm tagged}$ is found
to be $0.065 \pm 0.005$ in the electron and $0.069 \pm 0.005$ in the
muon channel, where the uncertainties include statistical and
systematic contributions. 
The ratio $k_{2\to\ge4}$ is found to be $2.52\pm 0.36$ in the electron channel and $2.35\pm 0.34$ in the muon channel.
The uncertainties include both systematic contributions and contributions arising from the limited
number of simulated events.
The total number of $W$+jets events passing
the selection with a $b$-tagged jet, $W_{\ge4 \rm ,tagged}$, is
evaluated to be $880 \pm 200 \ ({\rm stat.+syst.})$ in the electron
channel and $1390 \pm 310 \ ({\rm stat.+syst.})$ in the muon channel.

\subsection{Other backgrounds}

The numbers of background events coming from single top production, $Z$+jets and diboson events are 
evaluated using Monte Carlo simulation normalized to the 
relevant NNLO cross sections for $Z$+jets events and NLO for diboson events.

\subsection{Event yield}

The final numbers of expected and observed data events in both
channels after the full event selection are listed in
Table~\ref{evtnumbers}.  The number of events in the electron channel
is significantly lower than in the muon channel due to the higher
lepton \pt\ requirement and the more stringent missing momentum
requirement, which are necessary to reduce the
contribution from the multijet background. The overall agreement
between expectation and data is good.

\begin{table*}[htbp]
  \begin{center}
    {\footnotesize
  \begin{tabular}{| l | r r r | r r r | r r r | r r r |}
      \hline
      Channel & \multicolumn{3}{c|}{$\mu$ + jets pretag} & \multicolumn{3}{c|}{$\mu$ + jets tagged} & \multicolumn{3}{c|}{e + jets pretag} & \multicolumn{3}{c|}{e + jets tagged}            \\ \hline \hline
      $t\bar{t}$                 & 7200   & $\pm$  &    600     & 6300   & $\pm$  &    500   &  4800   & $\pm$  &    400     & 4260   & $\pm$  &    350   \\
     \hline
     W+jets                  & 8600  & $\pm$  &    1200    & 1390  & $\pm$  &    310  &  5400  & $\pm$  &    800    & 880  & $\pm$  &    200  \\
      Single top              & 460  & $\pm$  &    40    & 366  & $\pm$  &    32  &  320  & $\pm$  &    28    & 256  & $\pm$  &    22  \\ 
      Z+jets                  & 940  & $\pm$  &    330    & 134  & $\pm$  &    47  &  760  & $\pm$  &    270    & 110  & $\pm$  &    40  \\
      Diboson                 & 134 & $\pm$  &    7   & 22 & $\pm$  &    2 &  80 & $\pm$  &    5   & 13 & $\pm$  &    1 \\ \hline
      Multijets               & 1500     & $\pm$  &    800       & 500     & $\pm$  &    500     &  900     & $\pm$  &    500       & 250     & $\pm$  &    250     \\ \hline \hline
      Total background        & 11700   & $\pm$  &    1400     & 2400   & $\pm$  &    600   &  7500   & $\pm$  &    900 & 1500   & $\pm$  &    320   \\ \hline
      Signal + background     & 18900   & $\pm$  &    1600     & 8800   & $\pm$  &    800   &  12000   & $\pm$  &    1000     & 5800   & $\pm$  &    500   \\ \hline \hline
      Observed                & \multicolumn{3}{c|}{19639}     & \multicolumn{3}{c|}{9124} & \multicolumn{3}{c|}{12096}      & \multicolumn{3}{c|}{5829}
      \\ \hline
    \end{tabular}
  }    
  \end{center}
  \caption{Numbers of events observed in data and expected from $t\bar{t}$ signal events and various background processes
  for the pretag and tagged samples defined in Section~\ref{evselection}. The experimentally determined uncertainties quoted for $W$+jets 
  and multijet backgrounds include systematic uncertainties on the normalisation. 
  The quoted uncertainties on the other backgrounds are those from theory, taken to be 
  8\% for $t\bar{t}$ and single top, 34\% for $Z$+jets and 5\% for diboson backgrounds.
  The numbers correspond to an integrated luminosity of \lumi\ in both electron and muon channels.
  }
  \label{evtnumbers}
\end{table*}

\section{Reconstruction of the \ttbar\ final state}
\label{Reconstruction}
To measure the charge asymmetry in top pair events, the full \ttbar\ system is reconstructed. For this purpose, a kinematic fit
is used that assesses the compatibility of the observed event with the decays of a top-antitop pair based on a likelihood approach.

The likelihood takes as inputs the measured energies, pseudorapidities
and azimuthal angles of four jets, the measured energy of the lepton,
and the missing transverse momentum.  If there are more than four jets in the event satisfying $\pt>25\GeV$ and $|\eta|<2.5$, all 
subsets of four jets from the five jets in the event with highest $p_T$ are considered.

The likelihood is computed as
\begin{eqnarray}
   L & = & \mathcal{B}(\widetilde{E}_{\rm{p,}1}, \widetilde{E}_{\rm{p,}2} | m_W, \Gamma_W) 
   \cdot \mathcal{B}(\widetilde{E}_{\rm lep}, \widetilde{E}_{\nu} | m_W, \Gamma_W) \cdot \nonumber \\
              &   & \mathcal{B}(\widetilde{E}_{\rm{p,}1}, \widetilde{E}_{\rm{p,}2}, \widetilde{E}_{\rm{p,}3} | m_t, \Gamma_t) 
   \cdot \mathcal{B}(\widetilde{E}_{\rm lep}, \widetilde{E}_{\nu}, \widetilde{E}_{\rm{p,}4} | m_t, \Gamma_t) \cdot \nonumber \\
              &   & \mathcal{W}( \hat{E}_{x}^{\rm miss}| \widetilde{p}_{x, \nu}) \cdot \mathcal{W}(\hat{E}_{y}^{\rm miss} |
	   	      \widetilde{p}_{y, \nu}) \cdot \mathcal{W}(\hat{E}_{\rm{lep}} | \widetilde{E}_{\rm{lep}}) \cdot \nonumber \\
              &   & \nonumber \prod_{\rm{i=1}}^4 \mathcal{W}( \hat{E}_{\rm{jet,}i} | \widetilde{E}_{\rm{p,}i})
     \cdot \prod_{\rm{i=1}}^4 P(\textrm{tagged} ~| ~\textrm{parton flavour}) ,
    \label{eq:KLF_Likelihood}
\end{eqnarray}
where:
\begin{itemize}
  \item Symbols $\mathcal{B}$ represent Breit-Wigner functions, evaluated using
  invariant masses of sums of appropriate parton and lepton four-vectors. The pole masses of the $W$ boson and the top
  quark are fixed to $m_W = 80.4\,\GeV$ and $m_t = 172.5\,\GeV$, respectively. Their widths
  are taken to be $\Gamma_W = 2.1\,\GeV$ and $\Gamma_t = 1.5\,\GeV$.
  \item Symbols $\mathcal{W}$ represent the transfer functions associating the
  reconstructed quantities ($\hat{X}$) to quarks and leptons produced in the hard scattering ($\widetilde{X}$). 
  $\widetilde{E}_{\rm{p,}i}$ are the energies of partons associated
  to jets with measured energies $\hat{E}_{\rm{jet,}i}$.
  These transfer functions are derived from Monte Carlo simulation.
  \item $P(\textrm{tagged} ~| ~\textrm{parton flavour})$ is the 
  $b$-tagging probability or rejection efficiency, depending on the 
  parton flavour, as obtained from Monte Carlo simulation.
\end{itemize}

The likelihood is maximised with respect to the energies of the partons, the energy of the charged lepton,
and the components of the neutrino three-mo\-men\-tum. 
The assignment of jets to partons which gives the highest likelihood value is selected.
Finally, the sign of the charge of the top quark (or anti-quark) decaying into the lepton
is determined from the lepton charge.

The overall efficiency for the reconstruction of the correct event
topology is found to be 74\% in Monte Carlo simulated $t\bar{t}$ events.
Only those events where four jets and a lepton are matched 
to partonic particles are considered for the efficiency computation.

\begin{figure*}[htbp]
\begin{center}
\begin{minipage}{6in}
  \subfigure[]{
        \includegraphics[scale=0.37]{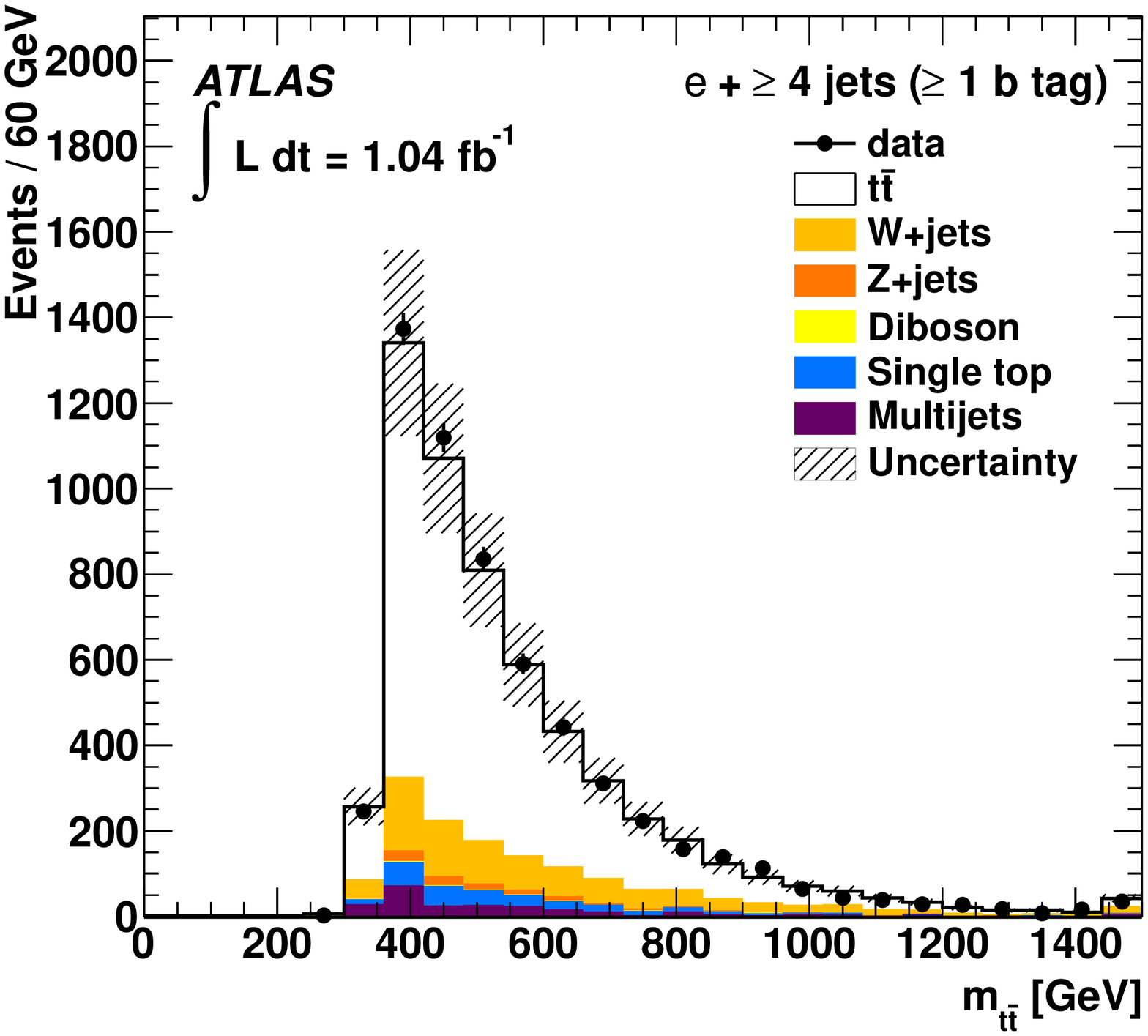}
       }
  \subfigure[]{
        \includegraphics[scale=0.37]{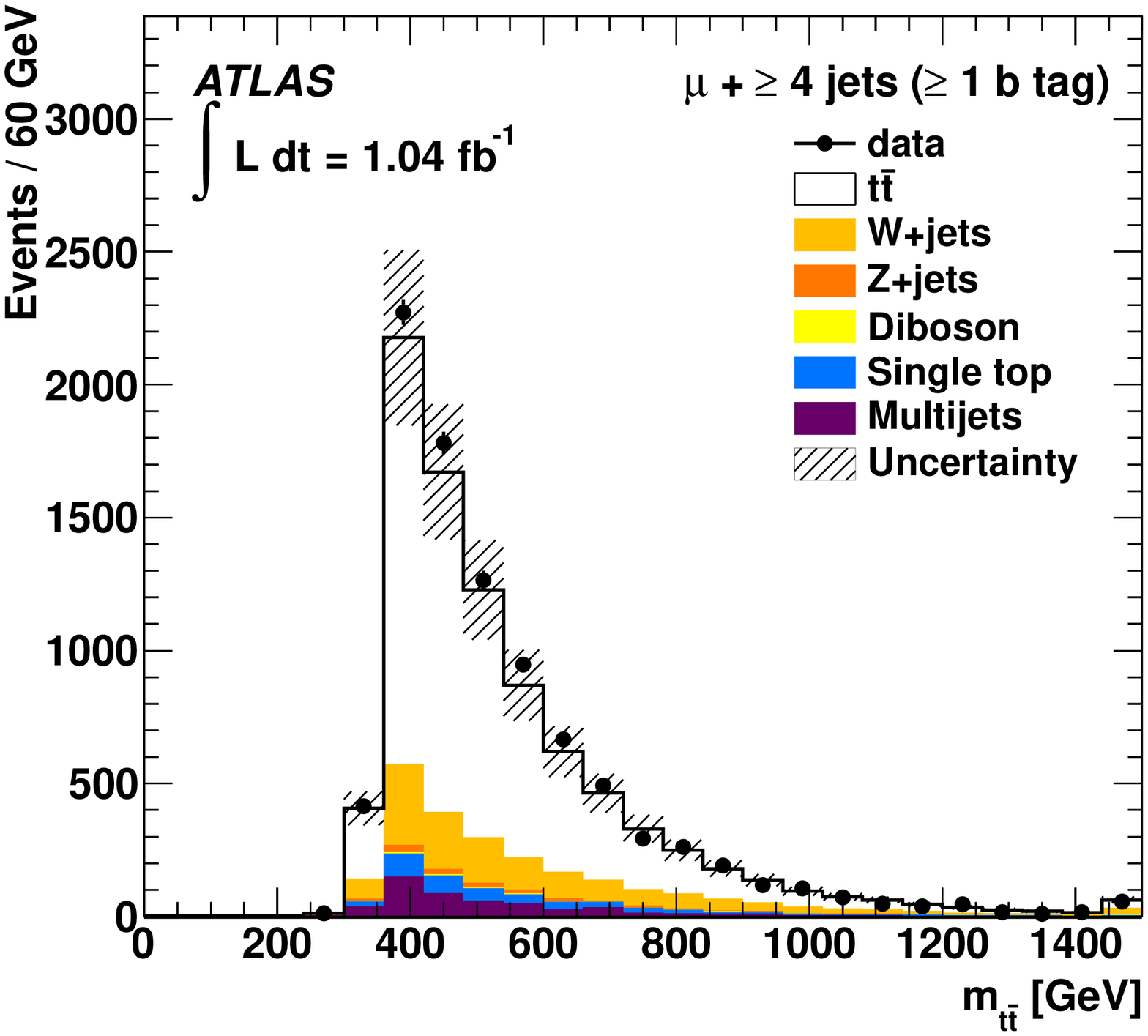}
        }
  \subfigure[]{
        \includegraphics[scale=0.37]{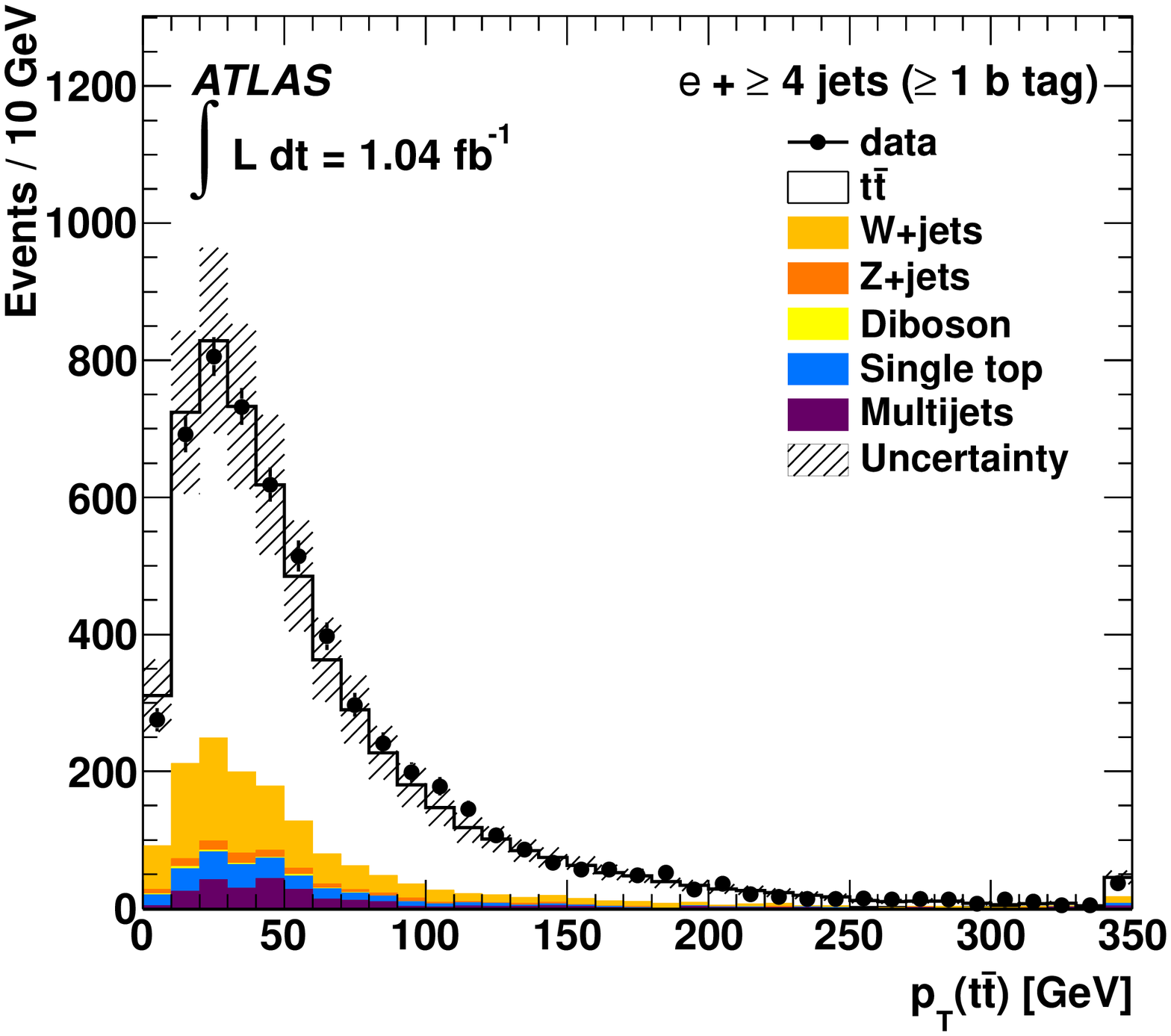}
       }
  \subfigure[]{
        \includegraphics[scale=0.37]{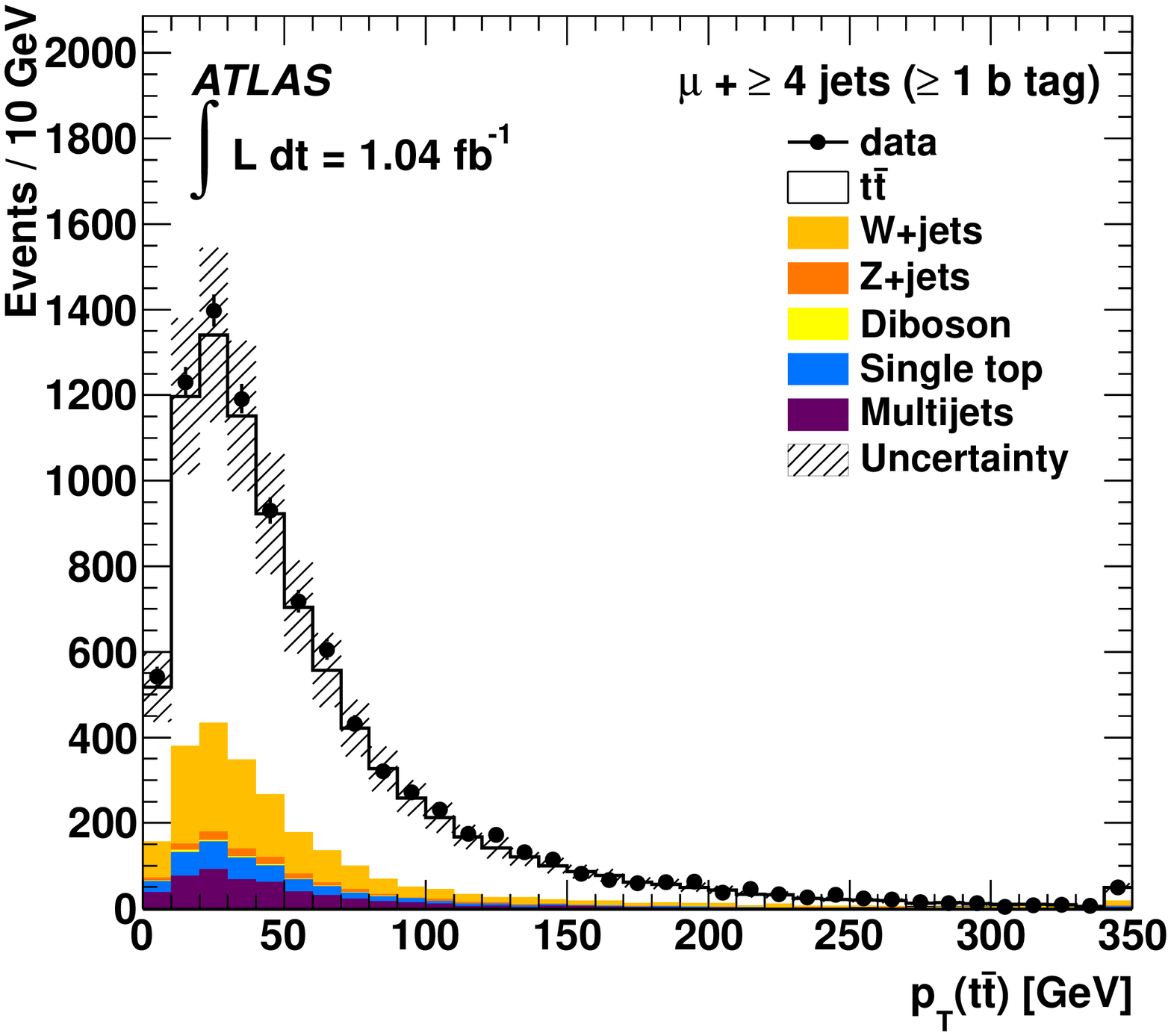}
        }
\end{minipage}
\end{center}
\caption{\label{fig:mpttbar}Expected and observed distributions for the invariant mass (plots (a) and (b)) and transverse 
momentum (plots (c) and (d)) of the reconstructed $t\bar{t}$ system. The left hand panels show distributions in the electron
channel, while the right hand panels show distributions in the muon channel.
The data are compared to the sum of the \ttbar\ signal contribution 
and backgrounds. The background contributions from $W$+jets and multijet production have been estimated 
from data, while the other backgrounds are estimated from simulation. The uncertainty on the combined signal and background 
estimate includes systematic contributions. Overflows
are shown in the highest bin of each histogram.}
\end{figure*}

Distributions of the invariant mass and transverse momentum of the reconstructed top-antitop
pair are shown in Fig.~\ref{fig:mpttbar}.

\section{Unfolding}
\label{Unfolding}
The measured distributions of top and anti-top rapidities are
distorted by detector effects and an event selection bias.
To correct for these distortions the experimental 
distributions are unfolded to the four-vectors of the top quarks before decay.

The relation between a true distribution $T_{j}$ (assuming, for simplicity, that there is
only one observable of interest) and the
reconstructed distribution $S_{i}$ after detector simulation and
event selection can be written:
\begin{equation}
S_{i} = \sum_{j}R_{ij}T_{j}
\end{equation}
where $R_{ij}$ is the response matrix defined
as the probability to observe an event in bin~$i$
when it is expected in bin~$j$.

The true distribution $T_{j}$ can be obtained from the
observed distribution $S_{i}$ by inverting the response matrix. 
The unfolding problem can similarly be formulated for the case of multiple observables.
In this analysis, Bayes' theorem is applied iteratively in order to
perform the unfolding~\cite{D'Agostini:1994zf}.

The unfolding is performed using response matrices which account for
both detector response and acceptance effects. The response matrices are calculated using Monte Carlo events
generated with \mcnlo. The unfolding is done separately, after background subtraction, for the inclusive measured distribution
of \DY (a one-dimensional unfolding problem), and the measured distribution \DY as a 
function of the reconstructed top-antitop invariant mass \mttbar\ (a two-dimensional
unfolding problem).

Two bins are used for \mttbar\ in the two-dimensional unfolding of \DY versus \mttbar, separated at $\mttbar = 450$~\GeV.
The choice of this \mttbar\ value is motivated by the observed CDF forward-backward asymmetry~\cite{CDF2} and by separating 
the data sample into two bins with roughly equal number of events. 

An additional cut on the value of the likelihood for the $t\bar{t}$ candidate is required in the two-dimensional unfolding,
since a large fraction of simulated events with a badly reconstructed \mttbar\ are found to have a low likelihood value.

\begin{figure*}[htbp]
\begin{center}
\begin{minipage}{6in}
  \subfigure[Response Matrix (e+jets)]{
  	\includegraphics[scale=0.35]{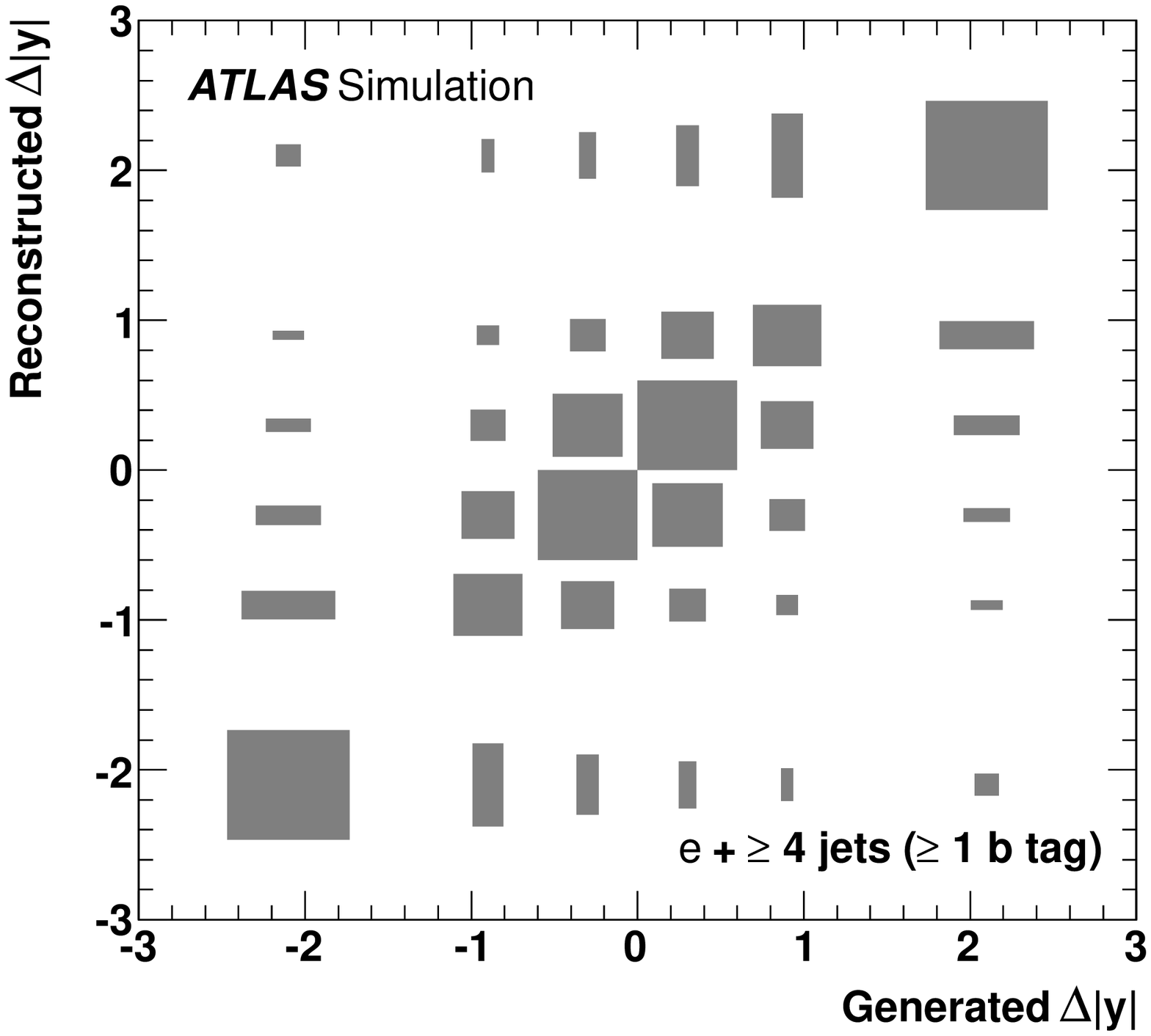}
  	}
  \subfigure[Response Matrix ($\mu$+jets)]{
    	\includegraphics[scale=0.35]{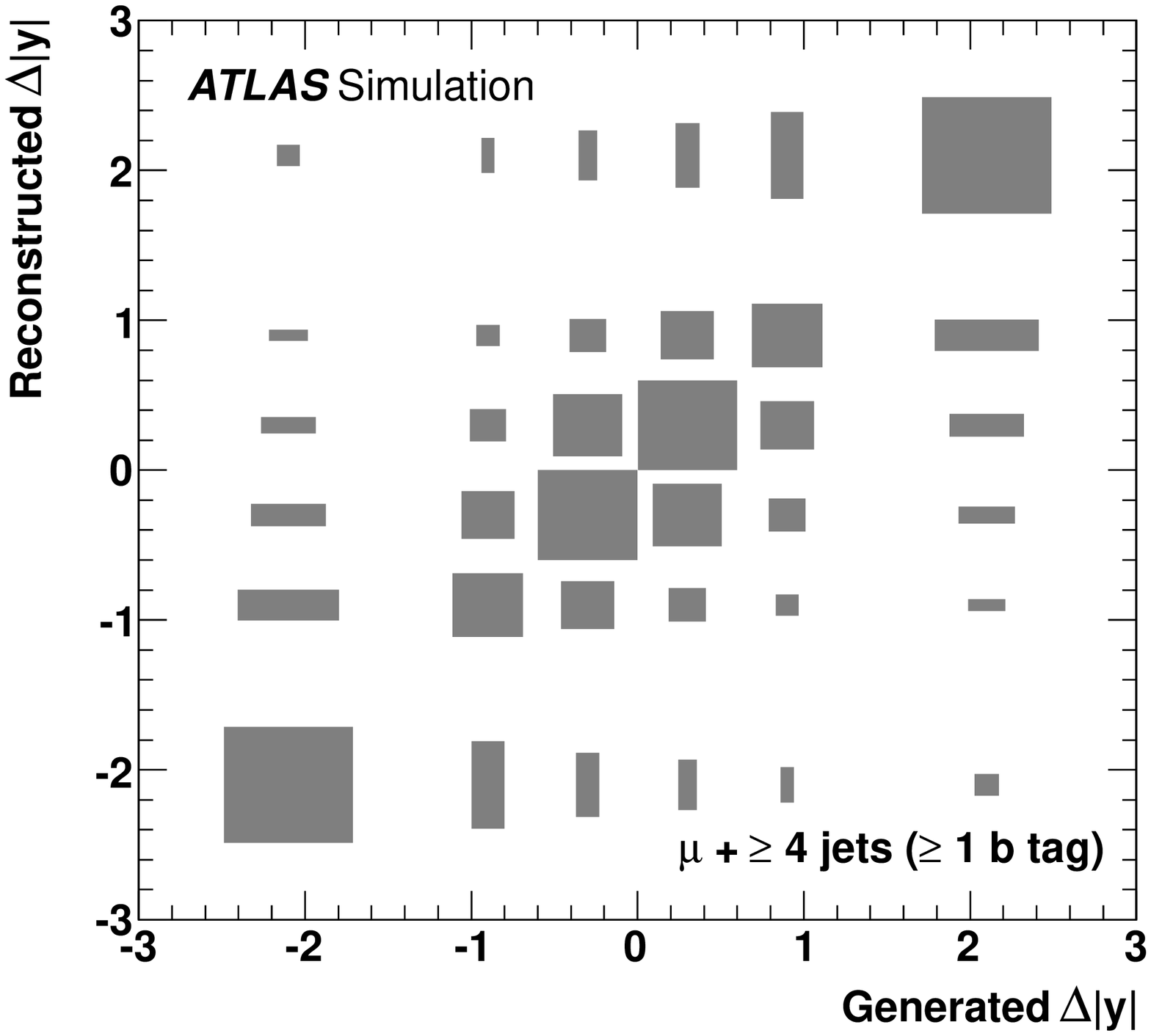}
  	}
\end{minipage}
\end{center}
\caption{\label{responsematrix}Correlations between the true and reconstructed values 
of \DY\ encoded in the unfolding response matrix for the electron (left) and muon (right) channels. The value of an entry in the matrix is proportional to the area of the corresponding box.}
\end{figure*}

The response matrix (including both detector and acceptance effects) for the inclusive \Ac\ measurement is shown in Fig.~\ref{responsematrix}. 
Six bins in $\Delta |y|$, in the range $-3<\Delta |y|<3$, are used in the response matrix, with the outermost bins broader 
than the inner bins in order to avoid the occurrence of bins with no entries in the measured distributions.
Only a very small fraction of simulated $t\bar{t}$ events are found to have $|\Delta|y||>3$, and hence such events have a negligible influence on the results.

The unfolding procedure is applied to the observed $\Delta |y|$ distribution in data, after subtracting 
background contributions. When performing the background subtraction, the shape of the multijet background
is obtained by applying the Matrix Method (described in Section~\ref{fakes}) in bins of $\Delta |y|.$
The shape of all remaining backgrounds is taken from Monte Carlo simulation.
The value of \Ac\ after unfolding is obtained by counting the numbers of events with $\Delta|y|>0$ and $\Delta|y|<0$
in the unfolded $\Delta|y|$ distribution.

\section{Systematic uncertainties}
\label{Systematics}
Several sources of systematic uncertainties are taken into account in
this analysis. These are categorised into the detector modelling,
the modelling of signal and background processes and the unfolding method.

\subsection{Detector modelling}

Small mis-modellings of muon or electron trigger, reconstruction and selection efficiencies in simulation are corrected for by scale factors derived from measurements of the efficiency in data. $Z\rightarrow\mu\mu$
or $Z\rightarrow e e$ and $W\rightarrow e\nu$ decays are used to obtain scale factors as functions of the lepton kinematics. The
uncertainties are evaluated by varying the lepton and signal selections and from the uncertainty in the evaluation of the
backgrounds.  Systematic uncertainties at the level of $1\%$ are found for both cases.
The same processes are used to measure the lepton momentum scale
and resolution. Scale factors, with uncertainties at the level of (1-1.5)\%, are derived to
match the simulation to observed distributions. A systematic
uncertainty for charge mis-identification of leptons is assigned
which is negligible for muons and ranges from 0.2\% to 3\% for
electrons depending on $|\eta|$.

The jet energy scale is derived using information from test-beam
data, collision data and simulation. Its uncertainty varies
between 2.5\% and 8\% in the central region, depending on jet \pt\ and
$\eta$~\cite{Aad:2011he}. This includes uncertainties in the
flavour composition of the sample and mis-measurements due to the effect of nearby
jets. Pile-up gives additional uncertainties of up to 5\%
(7\%) in the central (forward) region. An extra uncertainty of 0.8\% to 2.5\%,
depending on jet \pt,
is assigned to jets arising from the fragmentation of $b$-quarks, due to
differences between light and
gluon jets as opposed to jets containing $b$-hadrons. The jet
energy resolution and reconstruction efficiency are measured in data using
techniques described in Refs.~\cite{Aad:2011he,ATLAS-CONF-2010-054}, and their uncertainties
are found to be 10\% and (1-2)\%, respectively.

The $b$-tagging efficiencies and mis-tag rates are measured in
data. Jet \pt\ dependent scale factors, applied to
simulations to match the efficiencies measured in data, have
uncertainties which range from 9\%
to 15\% and 11\% to 22\%, respectively. 
A systematic uncertainty is assigned for a potential difference of up to 5\% between the $b$-tagging efficiency
for $b$-jets and that of $\bar{b}$-jets.
The uncertainty on the measured luminosity is 3.7\%~\cite{Aad:2011dr,ATLAS-CONF-2011-011}.  

Due to a hardware failure, later repaired, one small region of the liquid argon calorimeter could not be read out in 
a subset of the data corresponding to 84\% of the total integrated luminosity. Data events in which an electron or jet 
with $\pt>20$ \GeV\ is close to the affected calorimeter region are rejected for the relevant part of the dataset. Monte Carlo simulated
events with electrons or jets of $\pt>20$ \GeV\ close to the affected region are rejected with a probability equal to the
fraction of the integrated luminosity of data for which the calorimeter hardware problem was present. A systematic
uncertainty is evaluated by varying the \pt-threshold in data of the electrons and jets near the affected region by $\pm4$~\GeV, corresponding
to the uncertainty in the energy lost by objects in the affected region.

\subsection{Signal and background modelling}

The systematic uncertainty in the modelling of the signal process 
is assessed by simulations based on different
Monte Carlo generators. Sources of systematic uncertainty considered
here are the choice of generator and parton shower model, the choice
of parton density functions, the assumed top quark mass and the choice of
parameters which control the amount of initial and final state
radiation. Predictions from the \mcnlo\ and
\powheg~\cite{Nason:2004rx,Frixione:2007vw} generators are
compared. The parton showering is tested by comparing two
\powheg\ samples interfaced to \herwig\ and \pythia, respectively. The
amount of initial and final state radiation is varied by modifying
parameters in \acermc~\cite{AcerMC35} interfaced to \pythia\ according
to Ref.~\cite{CSCbook}. The parameters are varied in a range comparable to
those used in the Perugia Soft/Hard tune
variations~\cite{Skands:2010ak}. The impact of the choice of parton
density functions is studied using the procedure described
in Ref.~\cite{Botje:2011sn}. \mcnlo\ samples are generated assuming different
top quark masses and their predictions are compared. The observed
differences in the results are scaled to variations of $\pm0.9$~\GeV\
according to the uncertainty on the measured value~\cite{Lancaster:2011wr}.

As described in Section~\ref{Backgrounds}, background processes are
either modelled by simulation or estimated in auxiliary
measurements. The uncertainty in the estimate of the multijet background is evaluated by considering modified definitions of the loose data sample, taking
into account the statistical uncertainty in measuments of $\epsilon_{\rm real},
\epsilon_{\rm fake}$ described in Section~\ref{fakes} as well as the uncertainties in the normalisations of the $W$+jets and $Z$+jets backgrounds which are subtracted in the control region. The total uncertainty is estimated to be 100\%.
The normalisation of $W$+jets processes is
evaluated from auxiliary measurements using the asymmetric production
of positively and negatively charged $W$ bosons in $W$+jets
events. The uncertainty is estimated to be 21\% and 23\% in the four jet bin,
for the electron and muon channels respectively.
This uncertainty was estimated by evaluating the effect on both $r_{MC}$ and $k_{2\to\ge4}$ 
from the JES uncertainty and different PDF and generator choices.
Systematic uncertainties on
the shape of $W$+jets distributions are assigned based on differences
in simulated events generated with different simulation
parameters. Scaling factors correcting the fraction of heavy flavour
contributions in simulated $W$+jets samples are estimated in
auxiliary measurements, as described in Section~\ref{sec:ChAsymm}. The systematic
uncertainties are found by changing the normalisations of the non-$W$ processes
within their uncertainties when computing $W^{\rm Data}_{\rm i, pretag}, W^{\rm Data}_{\rm i, tagged}$,
as well as taking into account the impact of uncertainties in $b$-tagging efficiencies.
The total uncertainties are 47\% for $Wb\bar{b}$+jets and $Wc\bar{c}$+jets contributions and 32\% for
$Wc$+jets contributions. The normalisation of $Z$+jet events is
estimated using Berends-Giele-scaling~\cite{Berends199132}. The
uncertainty in the normalisation is 48\% in the four jet bin and
increases with the jet multiplicity. A systematic uncertainty in the shape
is accounted for by comparing simulated samples generated with
\alpgen\ and \sherpa~\cite{Gleisberg:2008ta}. The
uncertainty on the normalisation of the small background contributions
from single top and diboson production is estimated to be about 10\%
(depending on the channel) and 5\%, respectively.

Limited Monte Carlo sample sizes give rise to a systematic uncertainty
in the response matrix. This is accounted for by independently 
varying the bins of the response matrix according to Poisson distributions.

\subsection{Uncertainties from unfolding}
Closure tests are performed in order to check the validity of the unfolding procedure. Reweighted
$t\bar{t}$ samples with different amounts of asymmetry are considered. 
Pseudoexperiments are performed, varying the entries in histograms of the reconstructed distribution, to confirm that the 
response of the unfolding is linear in the true value of $A_C$ and that the true value of $A_C$ is recovered on average.
A total of 40 iterations are used in both channels for the inclusive $A_C$ measurement. For the measurement of 
$A_C$ as a function of $m_{t\bar{t}}$, 80 iterations are used. 
The number of iterations is chosen by ensuring that the unfolding procedure has converged in the sense that
the absolute change in the unfolded value of $A_C$ after performing an extra iteration is less than 0.001. It is found
that the unfolded values of $A_C$ from all pseudoexperiments and the data converge before the chosen numbers of iterations.
The potential bias arising from the choice of convergence criterion is taken into account by adding an additional
systematic uncertainty corresponding to the change in the unfolded value of $A_C$ obtained by further increasing the number
of iterations to very large values ($10^5$).

Pull distributions are constructed from pseudoexperiments and a relative shift of between 0\% and 10\% is found
in the unfolded value of $A_C$ with respect to the true value. An extra systematic uncertainty is assigned
to the unfolded value of $A_C$ obtained from data, corresponding to this shift.

In pseudoexperiments, a small bias is observed in the unfolded distributions corresponding to a relative
difference of a few percent between the unfolded result and true value in each bin. An additional
relative uncertainty of (2-5)\% is applied to all bins of the unfolded distributions, corresponding to the largest 
relative bin deviation observed in pseudoexperiments.

The statistical uncertainty in the unfolded measurement was computed using pseudoexperiments, propagating the uncertainties from 
the measured distribution using the statistical correlation matrix.

\subsection{Impact of systematic uncertainties}

The impact of the systematic uncertainties is evaluated by modifying the
subtracted background before unfolding and by modifying the response
matrix used for unfolding when relevant.
In particular the detector modelling systematic uncertainties are evaluated 
by shifting the estimated background as well as modifying the response matrix. 
Signal modelling uncertainties are computed by replacing the response matrix, and background modelling uncertainties by modifying the estimated background.

Table~\ref{Table_Systematics} summarises the sources of systematic
uncertainties for the inclusive measurement of the charge asymmetry, 
and their impact on the measured asymmetry, after unfolding. The systematics for the two $m_{t\bar{t}}$ bins are determined in a 
similar fashion.
The evaluation of some systematic uncertainties is limited by the finite size of the Monte Carlo samples. In these cases, the larger of the electron and muon channel uncertainties is used for the uncertainty on the combined result.
The resulting combined systematic uncertainties are $\pm0.028$ in the electron
channel and $\pm0.024$ in the muon channel.
\begin{table*}[htbp]
\begin{center}
\begin{tabular}{|l|r|r|}
\hline
Source of systematic uncertainty on $A_C$  & Electron channel & Muon channel \\
\hline
\multicolumn{3}{|l|}{\it Detector modelling} \\ 
\hline
Jet energy scale 			& 0.012 & 0.006 \\
Jet efficiency and resolution		& 0.001 & 0.007 \\ 
Muon efficiency and resolution		& $<$0.001 & 0.001 \\
Electron efficiency and resolution	& 0.003 & 0.001 \\
b-tag scale factors			& 0.004 & 0.002 \\
Calorimeter readout			& 0.001 & 0.004 \\
Charge mis-ID				& $<$0.001 & $<$0.001 \\
b-tag charge				& 0.001 & 0.001 \\
\hline
\multicolumn{3}{|l|}{\it Signal and background modelling} \\
\hline
Parton shower/fragmentation 		& 0.010 & 0.010 \\
Top mass 				& 0.007 & 0.007 \\
\ttbar\ modelling 			& 0.011 & 0.011 \\
ISR and FSR				& 0.010 & 0.010 \\
PDF					& $<$0.001 & $<$0.001 \\
W+jets normalization and shape		& 0.008 & 0.005 \\
Z+jets normalization and shape		& 0.005 & 0.001 \\
Multijet background 	 		& 0.011 & 0.001 \\
Single top 				& $<$0.001 & $<$0.001 \\
Diboson					& $<$0.001 & $<$0.001 \\
\hline
MC Statistics				& 0.006 & 0.005 \\
Unfolding convergence			& 0.005 & 0.007 \\
Unfolding bias				& 0.004 & $<$0.001 \\
Luminosity				& 0.001 & 0.001 \\
\hline \hline
Total systematic uncertainty 		& 0.028 & 0.024 \\
\hline
\end{tabular}
\end{center}
\caption{ List of sources of systematic uncertainties and their impact
  on the measured asymmetry in the electron and muon channel. In cases where asymmetric 
uncertainties were obtained, a symmetrisation of the uncertainties was performed by
taking the average of the absolute deviations under systematic shifts from the nominal value.}
\label{Table_Systematics}
\end{table*}

\section{Summary of Results}
\label{Results}
\begin{figure*}[!htb]
  \begin{center}
 \includegraphics[scale=0.43]{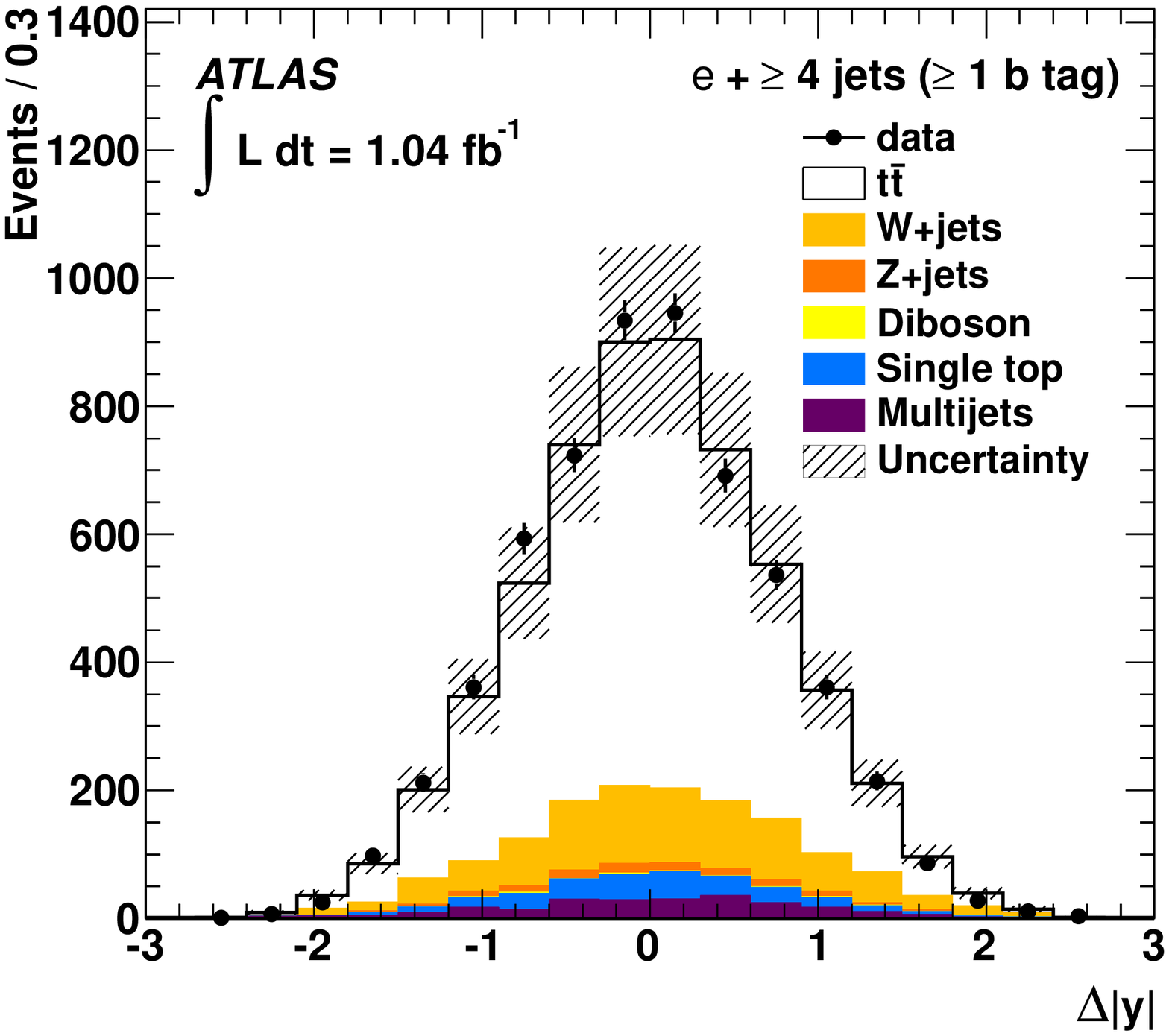}
 \includegraphics[scale=0.43]{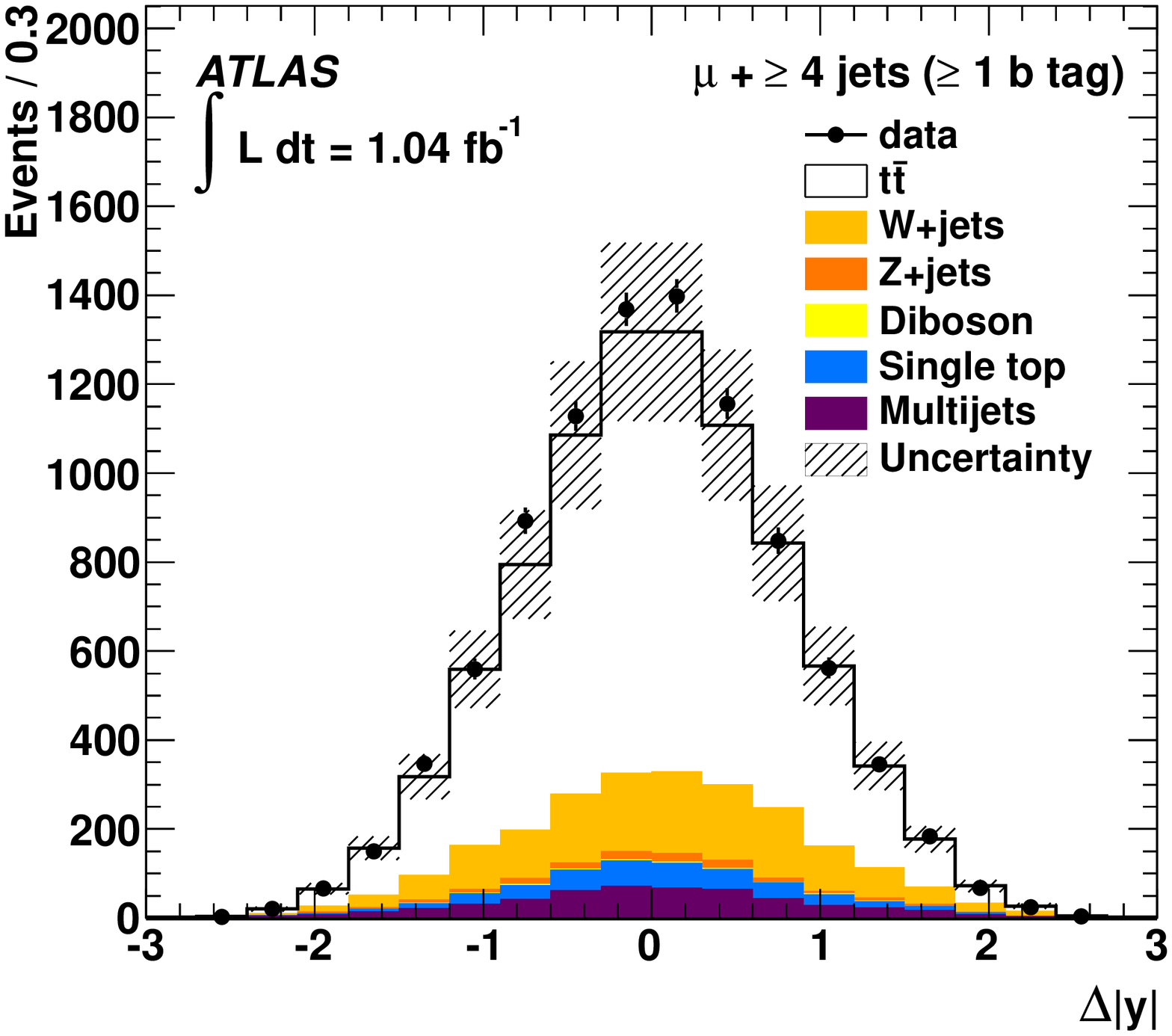}
    \caption{The measured $\Delta |y|$ distribution before unfolding for the electron channel (left) and for the muon channel (right) after $b$-tagging is applied.
Data (points) and Monte Carlo estimates (solid lines) are    represented. 
The multijet background and the normalisation of the $W$+jets background are obtained as explained in
    Section~\ref{Backgrounds}. The uncertainty on the combined signal and 
background estimate includes both statistical and systematic contributions.}
    \label{fig:DYtop}
  \end{center}  
\end{figure*}

\begin{figure*}[h!tb]
  \begin{centering}
\includegraphics[scale=0.37]{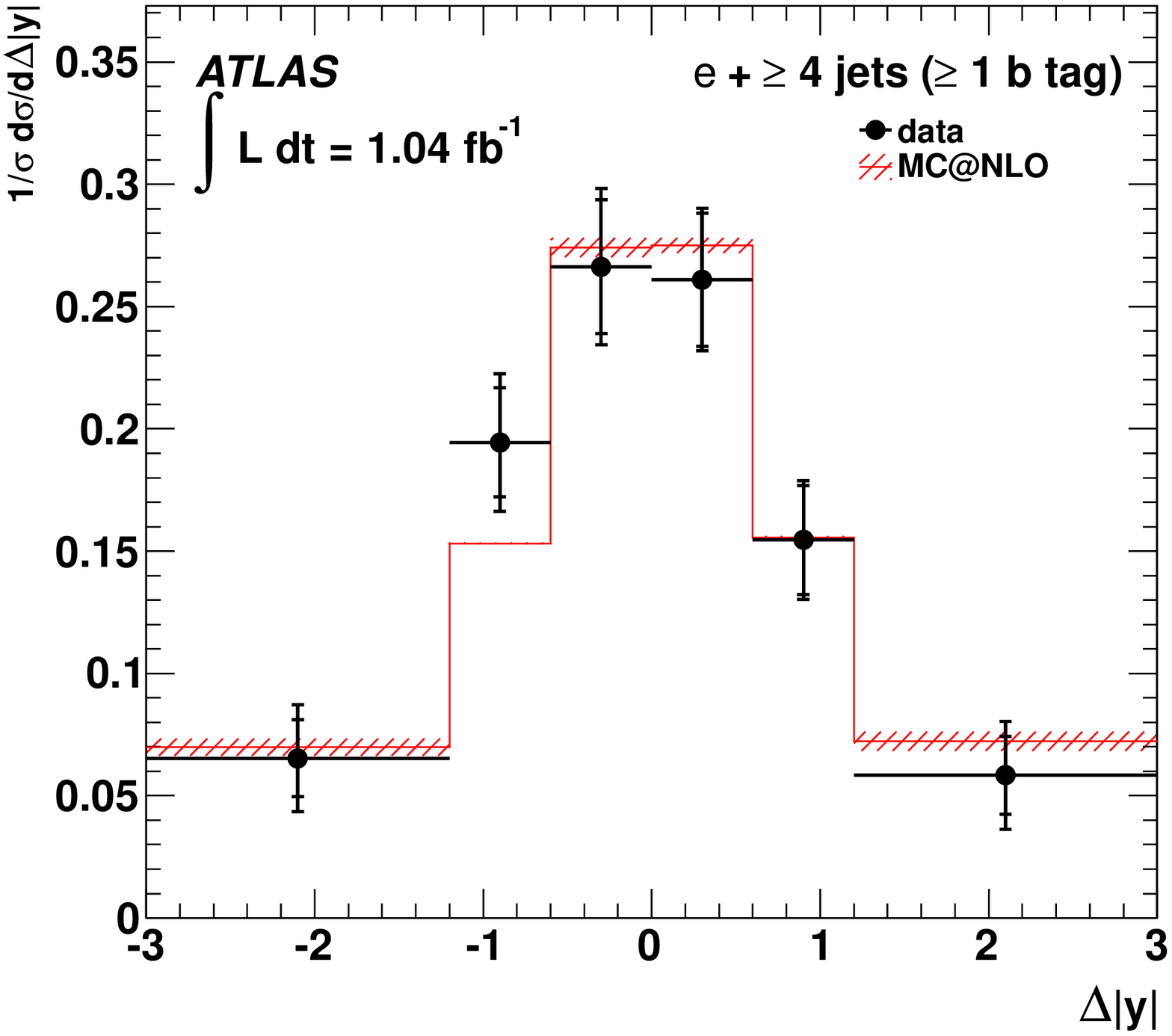}
\includegraphics[scale=0.37]{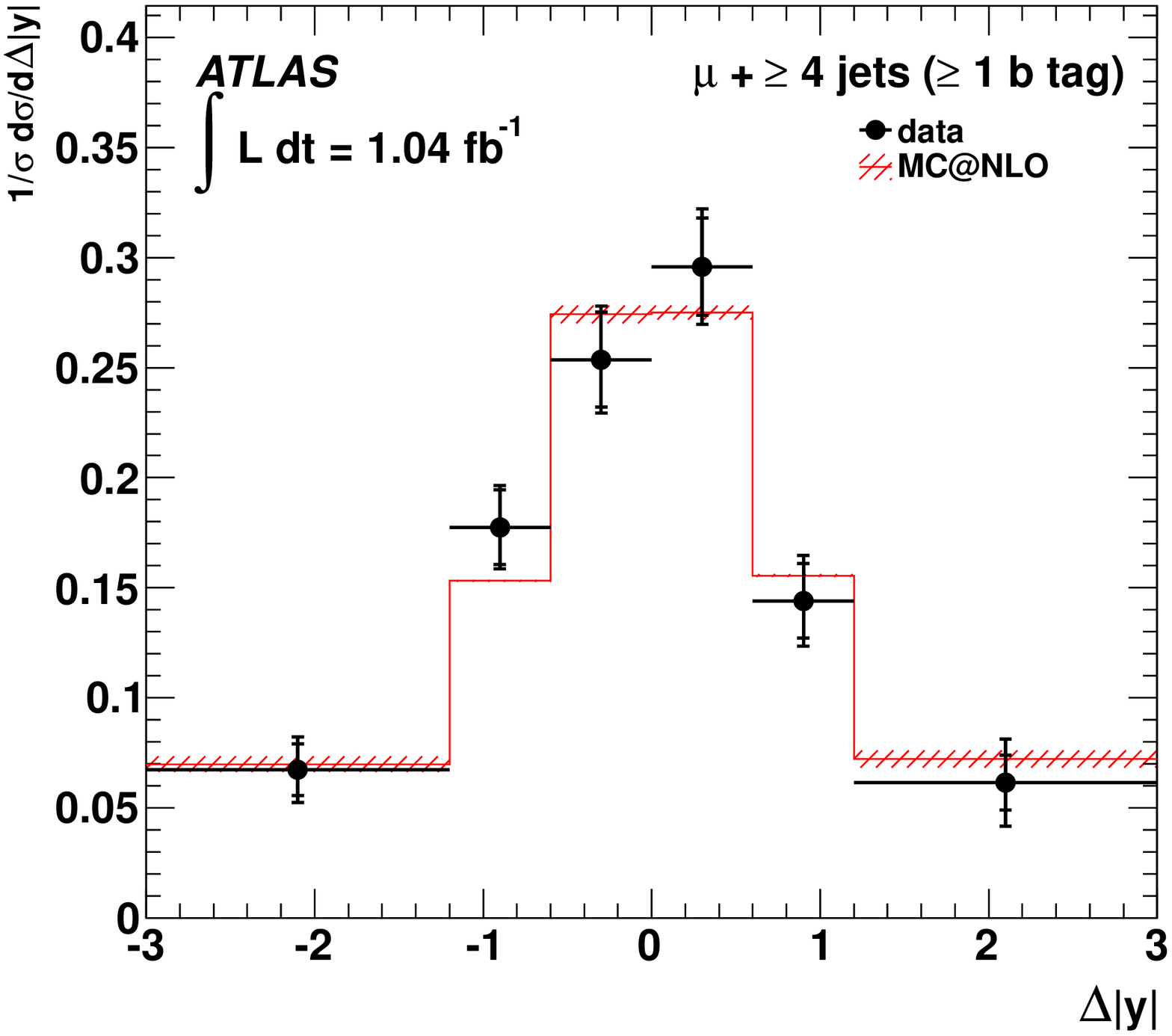}
\caption{The unfolded $\Delta |y|$ distribution for the electron channel (left) and the muon channel (right)
after $b$-tagging, compared to the prediction from \mcnlo. The uncertainties on the measurement include both statistical and systematic contributions,
which are shown separately. The inner part of the error bars corresponds to the statistical component of the uncertainty, while the outer part
corresponds to the systematic component.
The error bands on the \mcnlo\ prediction include uncertainties from parton distribution functions and renormalisation and factorisation scales.}
      \label{fig:unfolding:data}
  \end{centering}
\end{figure*}

The measured distributions of the top-antitop rapidity difference
$\Delta |y| = |y_t| - |y_{\bar{t}}|$ before unfolding are shown in
Fig.~\ref{fig:DYtop} for the electron and
muon channel. Fig.~\ref{fig:unfolding:data} shows
the corresponding \DY distributions after unfolding.
After unfolding, the bins of the measured distribution have statistical and systematic correlations. 
Adjacent bins of the $\Delta|y|$ distributions are found to be statistically anti-correlated with negative correlation 
coefficients of up to -0.6, whereas other correlations are small.

The measured values of the top charge asymmetry before and after 
unfolding, defined by Eq.~(\ref{Ac}) in terms of $\Delta|y|$, are summarised 
in Table~\ref{table:ChAsymm}.
The analytic best linear unbiased estimator (BLUE) method~\cite{Lyons:1988,Valassi:2003} is used to combine
the measurement in the electron and muon channels after correction for detector resolution and acceptance.

\begin{figure*}[!htbp]
  \begin{centering}
\includegraphics[scale=0.37]{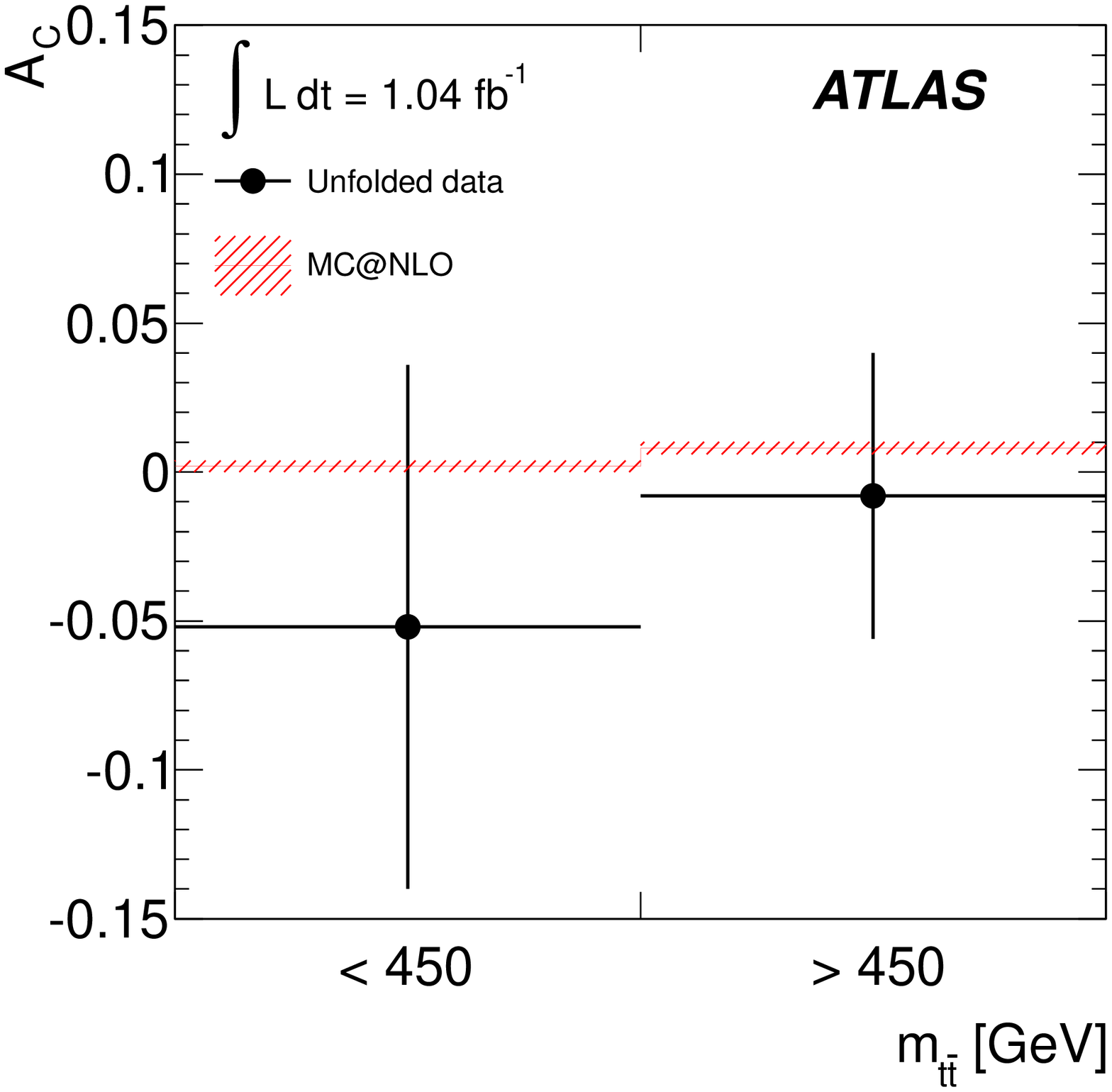}
\caption{Unfolded asymmetries in two regions of \mttbar\ compared to the prediction from \mcnlo. The error bands on the \mcnlo\ prediction include uncertainties from parton distribution functions and renormalisation and factorisation scales.} 
      \label{fig:AcMttbar}
  \end{centering}
\end{figure*}

The measured asymmetries are:
\begin{eqnarray*}
\resultincl
\nonumber
\end{eqnarray*}
for the integrated sample, and 
\begin{eqnarray*}
\Ac & = & -0.052 \pm 0.070\,\rm{(stat.)} \pm 0.054\,\rm{(syst.)} \\
& & {\rm for }\ \mttbar<450 \GeV, \\ 
\Ac & = & - 0.008 \pm 0.035\,\rm{(stat.)} \pm 0.032\,\rm{(syst.)} \\
& & {\rm for }\ \mttbar>450 \GeV.
\end{eqnarray*}
The measurement for the integrated sample can be compared with the result of the CMS Collaboration,
$A_\mathrm{C} = -0.013 \pm 0.028\;\mathrm{(stat)}^{+0.029}_{-0.031}\;\mathrm{(syst)}$~\cite{Chatrchyan:2011hk}.
Fig.~\ref{fig:AcMttbar} summarizes the measurements for the two \mttbar\ regions.
These results are compatible with the prediction from the \mcnlo\ Monte Carlo generator
of $\Ac = 0.006\pm 0.002$\footnote{The prediction of $0.0115\pm 0.0006$ for the charge asymmetry found in Ref.~\cite{Kuhn:2011ri} 
differs from the \mcnlo\ prediction of $0.006\pm 0.002$, due to the former
taking the LO prediction for the denominator in the definition~(\ref{Ac}) of $A_C$, and taking into account QED effects. The uncertainty on the \mcnlo\ 
prediction is obtained by considering variations in the renormalisation and factorisation scales and different sets of PDFs.},
showing no evidence for an enhancement from physics beyond the Standard Model.

\begin{table*}[!htbp] 
\begin{center} 
\begin{tabular}{|c|c|c|} 
\hline 
Asymmetry & reconstructed & detector and acceptance unfolded \\ 
\hline 
$A_C$ (electron) &  -0.034 $\pm$ 0.019 (stat.) $\pm$ 0.010 (syst.) & -0.047 $\pm$ 0.045 (stat.) $\pm$ 0.028 (syst.)\\ \hline
$A_C$ (muon)  & -0.010 $\pm$ 0.015 (stat.) $\pm$ 0.008 (syst.) & -0.002 $\pm$ 0.036 (stat.) $\pm$ 0.024 (syst.) \\ 
\hline 
Combined  &  &   -0.019 $\pm$ 0.028 (stat.) $\pm$ 0.024 (syst.) \\
\hline \end{tabular} 

\caption{\label{table:ChAsymm}The measured inclusive charge asymmetry values for the electron and muon channels after
background substraction, before and after unfolding. 
} 
\end{center}
\end{table*}

\section{Comparison of LHC and Tevatron Results}
\label{Interpretation}
The measurement of the charge asymmetry at the LHC is a test of the unexpectedly large forward-backward asymmetry observed at the 
Tevatron. However, because the LHC is a $pp$ collider and the centre of mass energy is around three times larger, any relation between 
the two asymmetries is model-dependent. Here a comparison is made between the predicted values of the Tevatron and LHC asymmetries 
for a few simple models beyond the SM. These are: (i) a flavour-changing $Z'$ boson with right-handed couplings, exchanged in the $t$ channel 
in $u \bar u \to t \bar t$~\cite{Jung:2009jz}; (ii) a $W'$ boson, also with right-handed couplings, contributing in 
$d \bar d \to t \bar t$~\cite{Cheung:2009ch}; a heavy axigluon $\mathcal{G}_\mu$ exchanged in the $s$ 
channel~\cite{Ferrario:2009bz,Frampton:2009rk}; (iv) a scalar doublet $\phi$, with the same quantum numbers as the SM 
Higgs~\cite{AguilarSaavedra:2011ug}; (v) a charge $4/3$ scalar, colour-sextet ($\Omega^4$) or colour-triplet ($\omega^4$), 
contributing in the $u$ channel to $u \bar u \to t \bar t$~\cite{Shu:2009xf,Dorsner:2009mq}. In all these models, the parameter 
space is described by the mass $M$ of the new particle (except for the axigluon which is assumed heavy, with $M \gg 7$ TeV) and a single coupling $g$.

In order to find the correlated predictions for the forward-backward and charge asymmetries in each model, a comprehensive scan 
over the mass $M$ and the coupling $g$ is performed using the {\sc Protos} generator~\cite{AguilarSaavedra:2008gt}, considering masses 
between 100~GeV and 10~TeV and the range of couplings for which the new physics contribution to the
$t\bar{t}$ cross section at the Tevatron lies in the interval [-0.8,1.7] pb. This is a conservative requirement 
which takes into account the different predictions for the SM cross section as well as the experimental measurement (see Ref.~\cite{AguilarSaavedra:2011hz} for 
details). 

\begin{figure*}[!htbp]
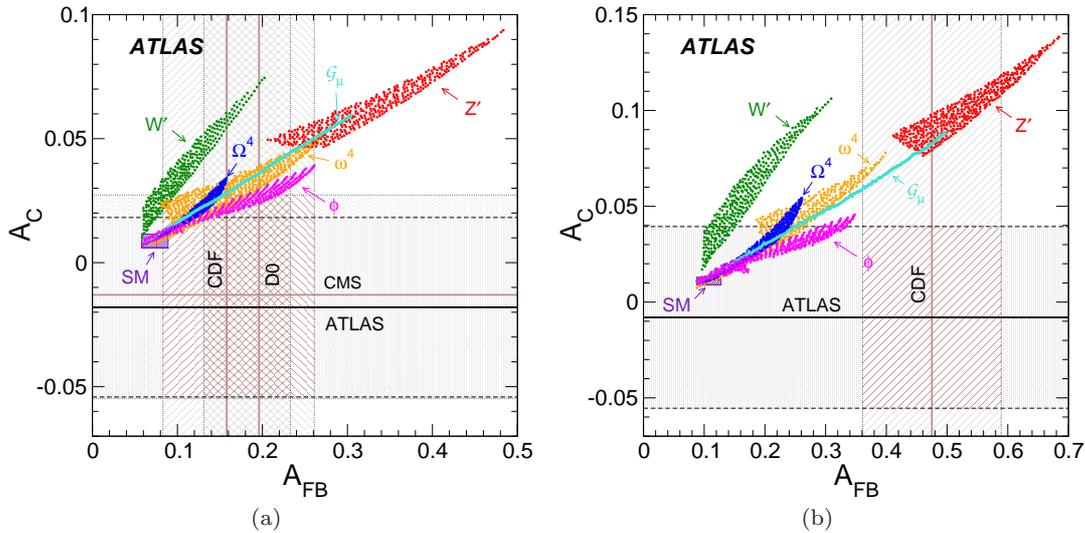

\vskip7mm
\begin{centering}
\subfigure[]{
\includegraphics[scale=0.37]{AFB-ATLAS.eps}
}
\subfigure[]{
\includegraphics[scale=0.37]{AFB-ATLAS450.eps}
}
\caption{
Measured FB asymmetries from the Tevatron and charge asymmetries from the LHC, compared to predictions from the SM as
well as predictions incorporating various potential new physics contributions.
The horizontal (vertical) bands and lines correspond to the ATLAS and CMS (CDF and D0) measurements.
In (a) the inclusive values are presented and in (b) the ATLAS measurement for $m_{t \bar t} > 450$ GeV is compared to the CDF measurement. 
The MC predictions for the new physics models are from Refs.~\cite{AguilarSaavedra:2011hz,AguilarSaavedra:2011ug}.
\label{fig:AvsA}}
  \end{centering}
\end{figure*}

In addition, a conservative upper limit on new physics 
contributions to $\sigma_{t\bar{t}}$ for $m_{t\bar{t}} > 1$ TeV is imposed. Further details can be found in 
Refs~\cite{AguilarSaavedra:2011hz,AguilarSaavedra:2011ug}. 
The coloured areas in Fig.~\ref{fig:AvsA} (a) represent the ranges of predicted values for the inclusive 
Tevatron forward-backward asymmetry, $A_{FB}$, and the inclusive LHC charge asymmetry, 
$A_C$, for the new physics models. The new physics contributions are computed 
using the tree-level SM amplitude plus the one(s) from the new particle(s). To a good approximation, the total asymmetries 
$A_{FB}$, $A_C$ are obtained from the former by summing the SM contribution (at NLO in the lowest order). 
The horizontal lines correspond to the present ATLAS measurement and the measurement reported by the CMS Collaboration \cite{Chatrchyan:2011hk}.
The vertical lines correspond to the asymmetry measurements at the Tevatron, $A_{FB} = 0.158 \pm 0.075$~\cite{CDF2} and 
$A_{FB} = 0.196 \pm 0.065$~\cite{D01}.

The ATLAS charge asymmetry measurement disfavours models with a new flavour-changing $Z'$ or $W'$ vector boson proposed to explain 
the measured Tevatron asymmetry. Minimal $Z'$ models are also excluded by 
the non-observation of same-sign top quark production \cite{:2012bb}.
For the other new physics models the asymmetries measured at the Tevatron are consistent with this measurement, within the experimental 
uncertainties.

Fig.~\ref{fig:AvsA} (b) shows the allowed regions for the high-mass asymmetries ($m_{t \bar t} > 450$ GeV) 
at the Tevatron and the LHC for the six new physics models. The vertical lines represent the CDF measurement $A_{FB} = 0.475 \pm 0.114$~\cite{CDF2},
while the horizontal lines correspond to the present ATLAS measurement.
In both panels of  Fig.~\ref{fig:AvsA}, the range of variation of SM predictions found in Refs~\cite{Hollik:2011ps,Ahrens:2011uf,Kuhn:2011ri} is indicated by a box.
The predictions of the six new physics models are in tension with the CDF and ATLAS high-mass measurements considered
together.

\section{Conclusion}
\label{Conclusion}
To summarize, the top quark charge asymmetry was measured in \ttbar\ events with a single lepton (electron or
muon), at least four jets and large missing transverse momentum using an integrated luminosity of \lumi\
recorded by the ATLAS experiment at a centre of mass energy of $\sqrt{s}$ = 7 TeV. The reconstruction of
\ttbar~events was performed using a kinematic fit. The reconstructed inclusive distribution
of \DY and the distribution as a function of $m_{t\bar{t}}$ were unfolded after background subtraction
to obtain results that can be directly compared with theoretical computations.
The results are compatible with the prediction from the \mcnlo\ Monte Carlo generator.
These measurements disfavour models with a new flavour-changing $Z'$ or $W'$ vector boson
that have been suggested to explain the measured Tevatron asymmetry.

\section{Acknowledgements}
\label{Acknowledgements}
We thank CERN for the very successful operation of the LHC, as well as the
support staff from our institutions without whom ATLAS could not be
operated efficiently.

We acknowledge the support of ANPCyT, Argentina; YerPhI, Armenia; ARC,
Australia; BMWF, Austria; ANAS, Azerbaijan; SSTC, Belarus; CNPq and FAPESP,
Brazil; NSERC, NRC and CFI, Canada; CERN; CONICYT, Chile; CAS, MOST and NSFC,
China; COLCIENCIAS, Colombia; MSMT CR, MPO CR and VSC CR, Czech Republic;
DNRF, DNSRC and Lundbeck Foundation, Denmark; EPLANET and ERC, European Union;
IN2P3-CNRS, CEA-DSM/IRFU, France; GNAS, Georgia; BMBF, DFG, HGF, MPG and AvH
Foundation, Germany; GSRT, Greece; ISF, MINERVA, GIF, DIP and Benoziyo Center,
Israel; INFN, Italy; MEXT and JSPS, Japan; CNRST, Morocco; FOM and NWO,
Netherlands; RCN, Norway; MNiSW, Poland; GRICES and FCT, Portugal; MERYS
(MECTS), Romania; MES of Russia and ROSATOM, Russian Federation; JINR; MSTD,
Serbia; MSSR, Slovakia; ARRS and MVZT, Slovenia; DST/NRF, South Africa;
MICINN, Spain; SRC and Wallenberg Foundation, Sweden; SER, SNSF and Cantons of
Bern and Geneva, Switzerland; NSC, Taiwan; TAEK, Turkey; STFC, the Royal
Society and Leverhulme Trust, United Kingdom; DOE and NSF, United States of
America.

The crucial computing support from all WLCG partners is acknowledged
gratefully, in particular from CERN and the ATLAS Tier-1 facilities at
TRIUMF (Canada), NDGF (Denmark, Norway, Sweden), CC-IN2P3 (France),
KIT/GridKA (Germany), INFN-CNAF (Italy), NL-T1 (Netherlands), PIC (Spain),
ASGC (Taiwan), RAL (UK) and BNL (USA) and in the Tier-2 facilities
worldwide.

\bibliographystyle{atlasnote}
\bibliography{TopChargeAsym_CONF,PHYSICS,ATLAS}

%
%

\clearpage
\onecolumn
\begin{flushleft}
{\Large The ATLAS Collaboration}

\bigskip

G.~Aad$^{\rm 48}$,
B.~Abbott$^{\rm 110}$,
J.~Abdallah$^{\rm 11}$,
A.A.~Abdelalim$^{\rm 49}$,
A.~Abdesselam$^{\rm 117}$,
O.~Abdinov$^{\rm 10}$,
B.~Abi$^{\rm 111}$,
M.~Abolins$^{\rm 87}$,
O.S.~AbouZeid$^{\rm 157}$,
H.~Abramowicz$^{\rm 152}$
% ...
\bigskip

% ...
$^{10}$ Institute of Physics, Azerbaijan Academy of Sciences, Baku, Azerbaijan\\
$^{11}$ Institut de F\'isica d'Altes Energies and Departament de F\'isica de la Universitat Aut\`onoma  de Barcelona and ICREA, Barcelona, Spain\\
% ...
$^{48}$ Fakult\"{a}t f\"{u}r Mathematik und Physik, Albert-Ludwigs-Universit\"{a}t, Freiburg i.Br., Germany\\
$^{49}$ Section de Physique, Universit\'e de Gen\`eve, Geneva, Switzerland\\
% ...
$^{87}$ Department of Physics and Astronomy, Michigan State University, East Lansing MI, United States of America\\
% ...
$^{110}$ Homer L. Dodge Department of Physics and Astronomy, University of Oklahoma, Norman OK, United States of America\\
$^{111}$ Department of Physics, Oklahoma State University, Stillwater OK, United States of America\\
% ...
$^{117}$ Department of Physics, Oxford University, Oxford, United Kingdom\\
% ...
$^{152}$ Raymond and Beverly Sackler School of Physics and Astronomy, Tel Aviv University, Tel Aviv, Israel\\
% ...
$^{157}$ Department of Physics, University of Toronto, Toronto ON, Canada\\
% ...
\end{flushleft}


\end{document}